\pdfoutput=1

\documentclass[twocolumn]{aastex62}

\newcommand{\code}[1]{\texttt{#1}}
\usepackage{pgffor}
\usepackage[caption=false]{subfig}
\usepackage{multirow}

\usepackage{amssymb}	
\usepackage{amstext}    

\graphicspath{{./}{figures/}}

\received{May 30, 2019}
\revised{July 15, 2019}
\accepted{July 17, 2019}
\submitjournal{ApJ}

\shorttitle{Molecular Gas in the Antlia Cluster}
\shortauthors{Cairns et al.}

\begin{document}

\title{Large Molecular Gas Reservoirs in Star Forming Cluster Galaxies}

\correspondingauthor{Joseph Cairns}
\email{j.cairns18@imperial.ac.uk}

\author{Joseph Cairns}
\affiliation{European Southern Observatory, Karl-Schwarzschild-Str., 85748, Garching, Germany}
\affiliation{Imperial College London, Blackett Laboratory, Prince Consort Road, London, SW7 2AZ, UK}

\author[0000-0001-8322-4162]{Andra Stroe}
\altaffiliation{ESO Fellow, Clay Fellow}
\affiliation{European Southern Observatory, Karl-Schwarzschild-Str., 85748, Garching, Germany}
\affiliation{Center for Astrophysics \text{\textbar} Harvard \& Smithsonian, 60 Garden St., Cambridge, MA 02138, USA}

\collaboration{}

\author[0000-0002-6637-3315]{Carlos De Breuck}
\affiliation{European Southern Observatory, Karl-Schwarzschild-Str., 85748, Garching, Germany}

\author[0000-0003-3816-5372]{Tony Mroczkowski}
\affiliation{European Southern Observatory, Karl-Schwarzschild-Str., 85748, Garching, Germany}

\author[0000-0002-9548-5033]{David Clements}
\affiliation{Imperial College London, Blackett Laboratory, Prince Consort Road, London, SW7 2AZ, UK}



\begin{abstract}
We present CO(2-1) observations of 72 galaxies in the nearby, disturbed Antlia galaxy cluster with the Atacama Pathfinder Experiment (APEX) telescope. The galaxies in our sample are selected to span a wide range of stellar masses ($10^{8}$\,M$_{\odot}\lesssim M_{\star} \lesssim 10^{10}$\,M$_{\odot}$) and star formation rates (0.0005\,M$_{\odot}$\,yr$^{-1}<\text{SFR}<0.3$\,M$_{\odot}$\,yr$^{-1}$). Reaching a depth of 23\,mJy in 50\,km\,s$^{-1}$ channels, we report a total CO detection rate of 37.5\% and a CO detection rate of 86\% for sources within 1 dex of the main sequence. We compare our sample with a similar sample of galaxies in the field, finding that, for a fixed stellar mass and SFR, galaxies in the Antlia cluster have comparable molecular gas reservoirs to field galaxies. We find that $\sim41\%$ (11/27) of our CO detections display non-Gaussian CO(2-1) emission line profiles, and a number of these sources display evidence of quenching in their optical images. We also find that the majority of our sample lie either just below, or far below the main sequence of field galaxies, further hinting at potential ongoing quenching. We conclude that the Antlia cluster represents an intermediate environment between fields and dense clusters, where the gentler intracluster medium (ICM) allows the cluster members to retain their reservoirs of molecular gas, but in which the disturbed ICM is just beginning to influence the member galaxies, resulting in high SFRs and possible ongoing quenching.
\end{abstract}

\keywords{galaxies: evolution -- galaxies: spiral -- galaxies: star formation -- galaxies: ISM -- ISM: molecules -- galaxies: clusters: individual: Antlia}

\section{Introduction}
\label{sec:intro}

As galaxies form and evolve over time, their properties are influenced by the environment in which they reside. In the local Universe, relaxed galaxy clusters are primarily comprised of quiescent, elliptical galaxies, whereas field environments contain many more star-forming, spiral galaxies \citep{1980ApJ...236..351D,1998ApJ...504L..75B,2003MNRAS.346..601G}. Observations \citep{1993AJ....106.1314Z,2006A&A...453..847C,2007AAS...20925203F} and simulations \citep{2003MNRAS.341.1333A,2005ApJ...622..853C} suggest that relaxed galaxy clusters constantly accrete small galaxy groups via filaments, and the motion of these infalling galaxies relative to the intracluster medium (ICM) can lead to the stripping of interstellar gas, known as ram pressure stripping \citep{1972ApJ...176....1G,1999MNRAS.308..947A,2014MNRAS.445.4335F}. High-energy, galaxy-galaxy close encounters and gravitational interactions between galaxies and the cluster potential can result in the distortion of cluster galaxies and the truncation of their haloes and disks, generally referred to as galaxy harassment \citep{1996Natur.379..613M,1998ApJ...495..139M,1998ApJ...509..587F}. Starvation occurs when the gas rich envelope of infalling galaxies is stripped, with the continued consumption of interstellar gas eventually exhausting the fuel for star formation \citep{1980ApJ...237..692L,2002ApJ...577..651B}. Viscous stripping tends to occur in galaxies with higher binding energies, with the viscous flow of the ICM around the galaxy resulting in the stripping of the interstellar medium (ISM) via Kelvin-Helmholtz instabilities \citep{LIVIO,1982MNRAS.198.1007N}. Thermal evaporation can also strip galaxies of their ISM through the transfer of heat from the ionised plasma of the ICM to the colder interstellar gas via thermal conduction by electrons \citep{1977ApJ...211..135C,1986RvMP...58....1S}.

As outlined above, the processes that quench star formation in relaxed clusters are relatively well understood. However, a significant fraction of galaxy clusters are in a disturbed state. New generation instruments have allowed large samples of galaxy clusters to be uniformly selected through the Sunyaev-Zel'dovich (SZ) effect. Recent independent studies found that, even at lower redshifts ($z<0.8$), about $70\%$ of {\it Planck} SZ-selected clusters are non-cool core (NCC), disturbed clusters \citep{2017ApJ...843...76A,2017MNRAS.468.1917R}. This fraction does not seem to evolve with redshift between $z=0$ and $z\sim0.8$ \citep{2017MNRAS.468.1917R}.

One of the main ways in which galaxy clusters grow is through mergers with other clusters. Galaxy cluster mergers are the most energetic events in the Universe, and as such, provide unique laboratories to connect large scale structure formation with processes occurring on (kilo-)parsec scales, such as particle acceleration in the ICM and star formation in member galaxies. As the most recently formed objects in the Universe, cluster mergers provide insight into the latest stage of cosmological structure formation \citep[e.g.][]{2005Natur.435..629S}. Disturbed clusters can act as contaminants and outliers for cosmological probes and scaling relations. Thus, understanding mergers is crucial for deriving precise cosmological parameters. Merging clusters have provided some of the most compelling evidence for the existence and nature of dark matter \citep[e.g.][]{2006ApJ...648L.109C,2012ApJ...747L..42D}. Galaxy cluster mergers can have a profound impact on the ICM: the merger drives significant turbulence in the ICM \citep{2003ApJ...584..190F, 2005MNRAS.357.1313C, 2009A&A...504...33V, 2011MNRAS.410..127B} and can release a large amount of energy into the ICM in the form of cluster-wide travelling shock waves \citep{2005ApJ...627..733M, 2006MNRAS.367..113P, 2010Sci...330..347V, 2011ApJ...728...82M}. 

The violent nature of disturbed clusters provides a vastly different environment for their member galaxies compared to relaxed clusters. While disturbed cluster environments are distinct from relaxed cluster environments, little research has been dedicated to understanding the evolution of their member galaxies. A handful of recent studies on the subject suggest that the environmental trends observed in relaxed clusters may be reversed in disturbed clusters. Simulations by \cite{2014MNRAS.443L.114R} suggest that travelling shocks induced by cluster mergers can enhance star formation for a few 100 Myr once the shock has interacted with the gas rich cluster members. \cite{2017MNRAS.465.2916S} find that, on average, merging clusters have higher densities of H$\alpha$ emitters than the surrounding fields, particularly for merging clusters displaying evidence of ICM shocks. High fractions of star-forming or very recently quenched disk galaxies, particularly located close to shock fronts, were also found in studies of individual merging clusters, such as: CIZA J2242.8+5301 \citep{2014MNRAS.438.1377S, 2015MNRAS.450..646S}, Abell 3921 \citep{2013A&A...557A..62P}, Abell 520 \citep{2017A&A...607A.131D}, Abell 521 \citep{2003A&A...399..813F, 2004ApJ...601..805U} and Abell 2744 \citep{2012ApJ...750L..23O}. Authors attribute these unexpectedly large populations of star forming galaxies to interactions between cluster members and the disturbed environment provided by either cluster-wide shocks from galaxy cluster mergers or accretion of young groups. \cite{2017MNRAS.472.3246M} demonstrated that the variation in average cluster colour decreases for more disturbed galaxy clusters, implying that star formation rates in disturbed clusters may be `standardised' by system-wide shocks. \cite{2017A&A...606A.108C} studied the morphologies of galaxies residing in both relaxed and disturbed clusters, finding that disturbed clusters are less efficient at converting star-forming spiral galaxies into gas-deficient lenticulars (S0 galaxies).

However, it remains unclear how the enhanced star formation activity in disturbed cluster environments is fuelled. It is therefore necessary to link observations tracing recent star formation to the reservoirs of cold, molecular gas within the member galaxies. Although the molecular gas mass in galaxies is dominated by H$_{2}$, the lowest rovibrational states of molecular hydrogen are forbidden and have high excitation requirements \citep{2013ARA&A..51..105C}. As a result, many studies use neutral hydrogen, H\textsc{i}, as a proxy for the molecular gas (e.g.\ \citealt{1988ApJ...333..136M, 1990AJ....100..604C, 2001ApJ...548...97S, 2009AJ....138.1741C, 2012MNRAS.422.1835S}). These studies find that spiral galaxies residing near the center of relaxed clusters are much more deficient in H\textsc{i} than those residing further from the cluster center. \cite{1994AJ....107.1003C} demonstrate that this deficiency is most likely due to stripping of the gas through interactions with the ICM. This trend, however, is not recovered for disturbed cluster environments. \cite{2015MNRAS.452.2731S} find that H$\alpha$ emission-line galaxies in the `Sausage' cluster contain comparable amounts of H\textsc{i} gas to their field counterparts around the cluster. Similarly, \cite{2012ApJ...756L..28J} find that many of the galaxies surrounding the central region of Abell 2192, a $z=0.19$ cluster in the process of forming, preserve their H\textsc{i} reservoirs. These findings suggest that gas-rich cluster members can retain their gas during cluster mergers, and that interaction between these gas-rich member galaxies and the travelling shock waves from the cluster merger can trigger star formation.

While observations of H\textsc{i} gas can provide useful insights into gas reservoirs, carbon monoxide (CO) rotational transitions are superior tracers of the cold, molecular gas which more directly fuels star formation (e.g.\ \citealt{2008AJ....136.2782L,2016MNRAS.459.3287C}) as CO is abundant, has low excitation requirements and its lower rotational transitions are easily observable from the ground \citep{2013ARA&A..51..105C}. A great deal of research probes molecular gas reservoirs via the CO rotational transitions. Much of the current literature, however, focuses on detecting CO rotational transitions in either low-$z$, relaxed clusters \citep{1986ApJ...301L..13K, 1991A&A...249..359C, 2002A&A...384...33B, 2009ApJ...693.1736W, 2012A&A...542A..32C, 2017ApJ...847..137K,2019MNRAS.483.2251Z}, Ultra-Luminous Infrared Galaxies (ULIRGS) and submillimetre galaxies \citep[see review by][]{2013ARA&A..51..105C}, or in the central regions of cooling flow clusters \citep[e.g.][]{2001MNRAS.328..762E, 2003ApJ...594L..13E, 2004A&A...415L...1S, 2006A&A...454..437S}. A handful of other studies detect CO rotational transitions in intermediate-$z$ \citep{2009MNRAS.395L..62G,2013A&A...557A.103J} and high-$z$ \citep{2018ApJ...856..118H} clusters, or in field environments \citep{2002ApJ...569..157W,2007MNRAS.377.1795C,2009ApJ...698L.178D,2011ApJ...730L..13B,2013ApJ...768...74T,2017A&A...604A..53C}. As a result, the molecular gas properties of star-forming galaxies residing in disturbed cluster environments remain unknown.

The Antlia cluster (Abell S636, RA=$10^\mathrm{h}30^\mathrm{m}03^\mathrm{s}$, DEC=$-35^{\circ}19^{'}24^{''}$) is perfectly suited to provide a statistical sample of molecular gas measurements in disturbed cluster spirals. It is the most nearby disturbed galaxy cluster, residing around 40 Mpc away ($z\sim0.009$), making CO rotational lines easily accessible. The cluster is relatively massive, with an estimated virial mass of around $5\times10^{14}$\,$\mathrm{M}_{\odot}$ \citep[e.g.][]{1985A&AS...61...93H, 2015MNRAS.452.1617H}. Antlia contains $\sim375$ galaxy members, with a galaxy density around 1.7 times higher than Virgo and 1.4 times higher than Fornax \citep{1990AJ....100....1F}. At its centre, the Antlia cluster hosts two main concentrations of galaxies surrounding the massive ellipticals NGC 3268 and NGC 3258, but contains at least five subclusters in total \citep{1985A&AS...61...93H}. The elongated galaxy and X-ray distribution of the Antlia cluster is indicative of a cluster in an intermediate merger state \citep[e.g.][]{2000PASJ...52..623N}. A number of previous studies have targeted the Antlia cluster members, primarily focussing on either the population of Early-Type Galaxies (ETGs) \citep{2008MNRAS.386.2311S, 2015MNRAS.451..791C}, dwarf galaxies \citep{2013MNRAS.430.1088C, 2014MNRAS.442..891C, 2014A&A...563A.118V} or the two central massive ellipticals and their surrounding systems \citep{2003A&A...408..929D, 2008MNRAS.386.1145B}. In their study of the H\textsc{i} content of the cluster members, \citet{2015MNRAS.452.1617H} detect 37 cluster members, out to a projected radius of 0.9\,Mpc. The velocity distribution of the H{\sc i} detections can be described as multi-modal, and is systematically redshifted from the optical velocity distribution by 700 km\,s$^{-1}$. At large radii, the H\textsc{i} rich galaxies are distributed asymmetrically, which the authors attribute to accretion along filaments.

In this study, observations of the CO $J=2-1$ rotational transition (hereafter known as the CO(2-1) transition) from the Atacama Pathfinder Experiment (APEX) telescope are used to trace the cold, molecular gas in the star-forming members of the Antlia cluster, in order to determine the effect of disturbed environments on the molecular gas content of star forming galaxies.

This paper is structured as follows: In Section \ref{sec:Ancillary_Data} we discuss the available multi-wavelength data for the Antlia cluster. In Section \ref{sec:sample} we outline the selection criteria for our sample and in Section \ref{sec:Data_reduction} we present the observing strategy and the methods used for data reduction. In Section \ref{sec:Analysis} we discuss our data analysis and we present our molecular gas measurements in Section \ref{sec:gas_detections}. In Section \ref{sec:Discussion} we provide a discussion of our results, by placing the detections in the context of the host galaxy properties and the larger scale cluster environment, and in Section \ref{sec:Conclusions} we summarise our main conclusions.

We assume a standard $\Lambda$CDM cosmology with $\mathrm{H}_0=67.3$\,km\,s$^{-1}$\,Mpc$^{-1}$, $\Omega_{\Lambda}=0.685$ and $\Omega_\mathrm{M}=0.315$ \citep{2014A&A...571A..16P}. At the average redshift of the Antlia cluster, $\langle z \rangle \approx 0.009$, 1$^{\prime\prime}$ corresponds to $0.192$\,kpc.

\begin{figure*}
	\includegraphics[width=\textwidth]{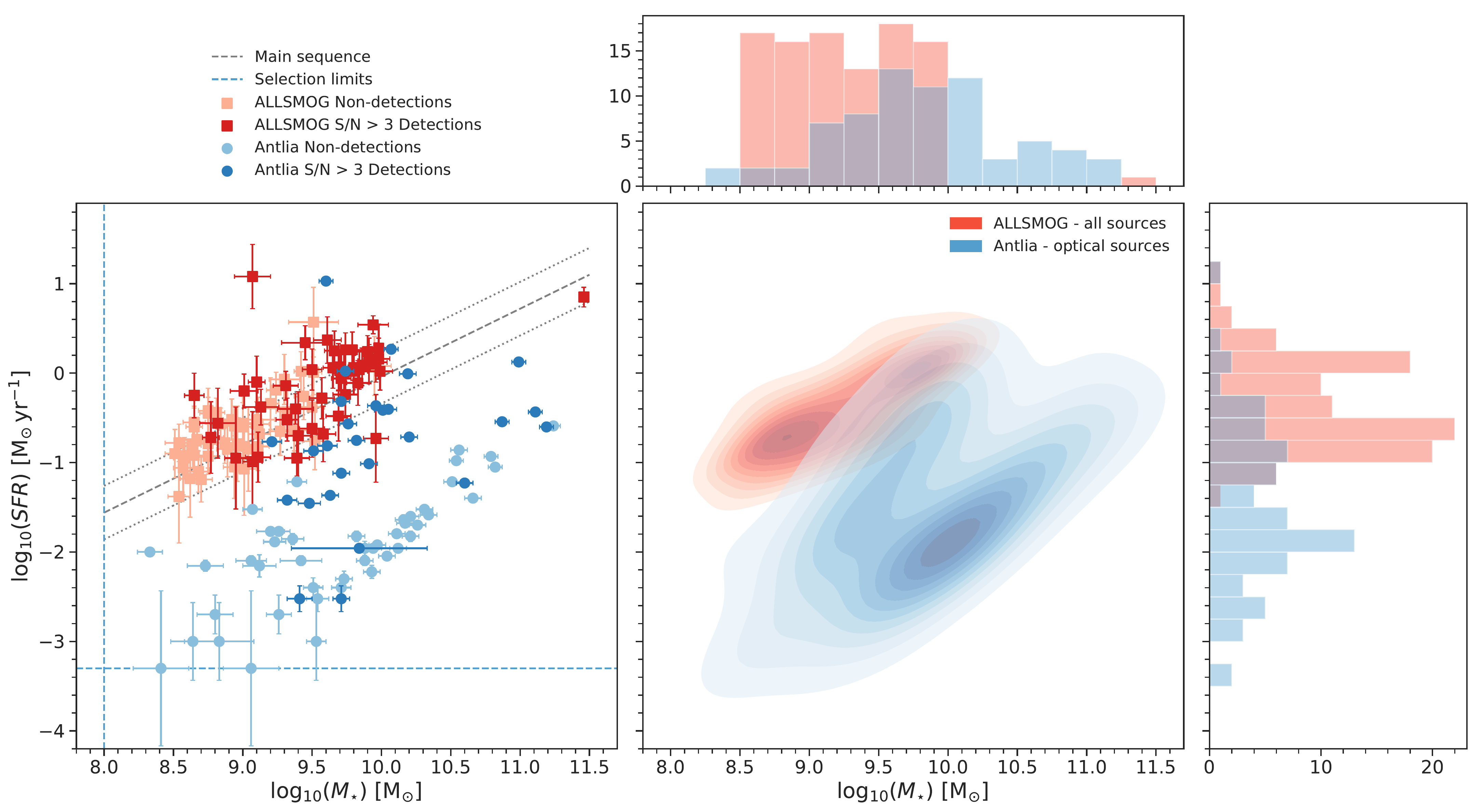}
    \caption{The distribution of our optically-selected Antlia cluster members in the M$_{\star}$-SFR plane. We separate between APEX CO(2-1) detections and non-detections. Leveraging the extensive spectroscopy available for Antlia, we selected cluster members down to a SFR of 0.0005\,M$_{\odot}$/yr and a stellar mass of $10^8$\,M$_{\odot}$. We overplot the local main sequence (MS) from \citep{2015ApJ...801L..29R}. For comparison, we show the ALLSMOG sample of field star-forming galaxies at a similar redshift \citep{2017A&A...604A..53C}. Note that while ALLSMOG specifically targeted galaxies on the MS, our selection was broader and includes galaxies below the MS.}
    \label{fig:smass_sfr}
\end{figure*}

\section{Ancillary Data}
\label{sec:Ancillary_Data}

As discussed above, the Antlia cluster benefits from a wealth of multiwavelength data, including optical spectroscopy, multi-band mid-infrared observations and blind H{\sc i} observations.

\subsection{Optical imaging}
We use visual evaluation of the available optical imaging from both ground and space observations to perform the morphological classification of the cluster galaxies. The deepest multi-band optical data to cover the entire Antlia cluster region is the 
Digitized Sky Survey\footnote{\url{https://archive.stsci.edu/dss/}}. The cluster galaxies are well detected in the three filters (blue, red and infrared), which we use to make RGB images of the cluster with the help of the {\sc Aladin} software\footnote{\url{http://aladin.u-strasbg.fr/}} \citep{2000A&AS..143...33B, 2014ASPC..485..277B}.

We use the ESO Archive Science Portal\footnote{\url{http://archive.eso.org/scienceportal/home}} to explore the optical data available for the cluster. We gather VIMOS r-band images (obtained as part of ESO programmes 079.B-0480(A), PI Richtler and 093.B-0148(A), PI D Rijcke) for 33 of the sources studied in this paper. Four sources also have {\it Hubble} Space Telescope observations (programmes 14920, PI Boizelle; 7919, PI Sparks; 9427; PI Harris), downloaded through the Mikulski Archive for Space Telescopes\footnote{\url{https://archive.stsci.edu/}}.

\subsection{Velocities}
We use the NASA/IPAC Extragalactic Database\footnote{\url{https://ned.ipac.caltech.edu/}} to collect redshifts for sources located within a radius of 1.5\,deg from the position of the Antlia core. In line with \citet{2015MNRAS.452.1617H}, we consider a source to belong to the cluster if its velocity lies in the $1200-4200$ km\,s$^{-1}$ range, or $z=0.004-0.014$. Given the velocity dispersion $\sigma\sim525$ km\,s$^{-1}$ of the cluster measured from optical galaxies, the selection corresponds to about $3\times\sigma$ on either side of the distribution peak. \cite{2008MNRAS.386.2311S} discuss the justification for this value, concluding that a more relaxed membership criteria is preferred due to the substructure of the Antlia cluster and its likely disturbed and complex dynamics. 

\subsection{Mid-infrared data, stellar masses and star formation rates}
We use estimations of stellar mass and star-formation rate (SFR) for the galaxies in the Antlia cluster, as derived by \citet{2015MNRAS.452.1617H} using mid-infrared data from the {\it Widefield Infrared Survey Explorer} ({\it WISE}). \citet{2015MNRAS.452.1617H} build their catalogue by first using the most sensitive band -- W1 (3.4~$\mu$m) -- for detection of extended sources, and then measuring their magnitudes in {\it WISE} bands W2, W3 and W4 (respectively 4.6, 12, and 22~$\mu$m) for these sources down to a signal to noise $S/N>3$.\footnote{\url{http://wise2.ipac.caltech.edu/docs/release/allsky/}} 

As discussed in \citet{2015MNRAS.452.1617H}, the $3.4-4.6$\,$\mu$m colour (W1-W2) can be used to estimate the stellar mass in nearby galaxies. The 12\,$\mu$m (W3) band is a simple estimator of star formation (SF), which correlates with well-studied, reliable SFR estimators, such as H$\alpha$ emission. The 12$\mu$m is particularly reliable for high SFRs ($\sim10$\,M$_{\odot}$\,yr$^{-1}$, see e.g.\ \citealt{2012ApJ...748...80D}), with an increasingly larger scatter with respect to H$\alpha$ for smaller SFRs.

\subsection{H{\sc i} data and neutral gas masses}
\label{sec:Ancillary_Data:HI}
\citet{2015MNRAS.452.1617H} observed the Antlia cluster with the seven dish Karoo Array Telescope (KAT-7), in a radio mosaic covering a total area of about $4.4$\,deg$^2$ at a resolution of $\sim3.2$\,arcmin. \citet{2015MNRAS.452.1617H} obtain $37$ H{\sc i} detections associated with the cluster. The authors note that 33 of these sources have an optical counterpart within a $2$\,arcmin radius, while only 18 sources have a galaxy host located within $0.5$\,arcmin. Given the resolution of the H{\sc i} and the signal-to-noise threshold $S/N>3$ imposed by \citet{2015MNRAS.452.1617H}, we can estimate the positional uncertainty to be about $1.1^{^\prime}$. This could be driven by the large positional uncertainty caused by the poor resolution of the H{\sc i} data, or could indicate real  H{\sc i} offsets, caused by ram pressure stripping.

\section{The Sample Selection}
\label{sec:sample}

Leveraging the excellent multiwavelength data available (discussed in Section~\ref{sec:Ancillary_Data}), we pursued two complementary avenues for studying the cold molecular gas reservoirs of Antlia cluster galaxies. 

The first selection is focused on obtaining a uniform sampling of galaxies down to a limiting stellar mass and SFR, with a range of morphological types. We supplement this selection with a sample focused on galaxies with large reservoirs of neutral gas, down to a limiting H{\sc i} mass of $10^{8.2}$\,M$_{\odot}$. However, given the large positional uncertainty of the H{\sc i} observations, we believe that no conclusions can be drawn from observations of the H{\sc i} selected sample. For completeness we present the H{\sc i} selection and the results in Appendix~\ref{sec:appendix:HI}, however we do not further discuss it in the main body of the text.

\begin{figure}
	\includegraphics[width=\columnwidth]{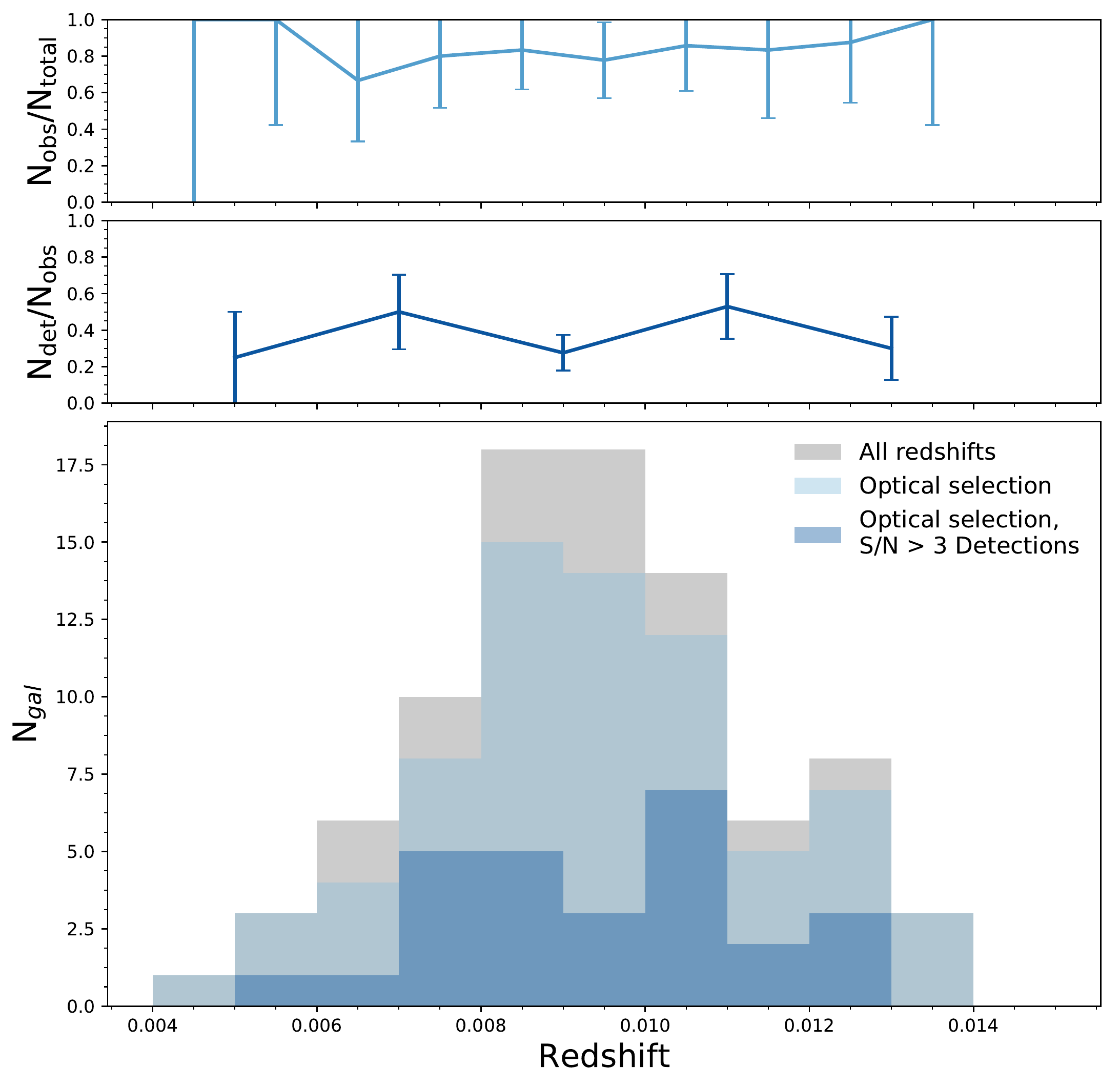}
    \caption{Redshift distribution of the Antlia cluster sources. We overlay the sources that were followed up as part of our optical selection. The fraction of selected sources slightly rises towards the outskirts of the cluster because of the increasing fraction of star-forming galaxies. The rate of detection is flat with redshift.}
    \label{fig:redshift_histogram}
\end{figure}


Since our goal is to understand how SF is fuelled in cluster galaxies, our first aim is to obtain a uniform sampling of galaxies covering a large fraction of the stellar-mass-SFR plane. This enables us to correlate major physical parameters tracing the current growth rate (i.e.\ SFR) and integrated evolution over cosmic time (i.e.\ stellar mass) of a galaxy with its reservoirs fuelling future SF (through molecular gas). The selection criteria for the sample henceforth referred to as the stellar-mass selected sample are:
\begin{itemize}
    \item detected in the W1 band {\it WISE} data as an extended source
    \item projected radius from the cluster core $r<1.5^\circ$ (equivalent to $\sim1$\,Mpc)
    \item velocity within the $1200-4200$ km\,s$^{-1}$ range
    \item stellar mass $M\star>10^8$\,M$_\odot$
    \item SFR measured from the 12\,$\mu$m emission $SFR>0.0005$\,M$_\odot$\,yr$^{-1}$
\end{itemize}

This selection results in a sample of 72 galaxies, which effectively cover main-sequence star forming galaxies down to relatively quiescent, as well as possibly irregular, galaxies. There are a further 12 galaxies that are detected in {\it WISE} and have redshifts, but which do not make our selection either because of their slightly smaller stellar mass or their minimal SFR. There are about 150 more sources with spectroscopy, which do not have {\it WISE} counterparts at all. There are most likely faint sources with low masses below our selection limit.

The stellar masses of the galaxies selected for follow-up range between $\sim2\times10^8$\,M$_\odot$ and $\sim1.75\times10^{10}$\,M$_\odot$, probing ranges similar to state-of-the-art field surveys at a similar redshift \citep[e.g.][]{2017A&A...604A..53C}. The SFR ranges between 0.0005 and 1.3\,M$_\odot$\,yr$^{-1}$, with an average at 0.3\,M$_\odot$\,yr$^{-1}$. This includes galaxies on the main-sequence for field galaxies at this redshift, as well as galaxies located below it \citep[e.g.][]{2017A&A...604A..53C}. The distribution of galaxies in the stellar mass-SFR plane can be found in Figure~\ref{fig:smass_sfr}. In terms of redshift, the selection is relatively flat with respect to the total number of galaxies, slightly rising at large velocities, where the fraction of star-forming galaxies increases (see Figure~\ref{fig:redshift_histogram}).

\section{Data Acquisition and Reduction}
\label{sec:Data_reduction}

\subsection{Observations and Data}
\label{sec:Data_reduction:Observations}

We observed the 72 optically selected sources as presented in Section~\ref{sec:sample} with the PI230 instrument at the Nasmyth-B focus of the the Atacama Pathfinder EXperiment (APEX) telescope\footnote{\url{http://www.apex-telescope.org/}}. PI230 is a dual polarisation receiver, covering the $200-270$ GHz frequency range at $\sim30^{\prime\prime}$ resolution with an intrinsic spectral resolution of 0.0796\,km\,s$^{-1}$. As a result, PI230 provides exquisite sensitivity: compared to the old SHFI-1 on APEX, a factor of $>2$ better depth can be obtained in the same observing time. For the most extended sources, we cover at least $85$ per cent of the flux given the beam of APEX, while the smaller sources are fully covered.

Our PI230 observations were designed to obtain detections for sources with stellar masses larger than $\sim10^{9.0}$\,M$_\odot$ and significant upper limits for sources with smaller stellar masses. 

We based our observational plan on the APEX Low-redshift Legacy Survey for Molecular Gas \citep[ALLSMOG,][]{2017A&A...604A..53C}, since it was performed on a similar samples of galaxies at $z\sim0.01-0.02$, but selected in the field. Using the ALLSMOG results, we expected, given the stellar masses of our targets, that they will have CO(2-1) luminosities of $\sim10^{7.5}$\,K\,km\,s$^{-1}$\,pc$^{-2}$. Assuming an average velocity width of the gas distribution of $\sim150$\,km\,s$^{-1}$ and a top-hat flux density distribution, we target a signal-to-noise of $S/N>5$ in $10$\,km\,s$^{-1}$ wide channels, equivalent to an RMS noise of $\sim1.3$ mK per channel. Each science observation was preceded by calibration scans to focus the telescope and correct the pointing. For the science observations, we employed a `wobbler switching' observing technique, where the wobbler position is swapped between the science target and a sky position, with an amplitude of $60''$. This enables an accurate sky subtraction, thus obtaining stable baselines. For the science targets, the telescope position was chosen according to the optical coordinates of the stellar mass-SFR selected sources. The tuning frequency was set to the expected CO(2-1) frequency for each source, as calculated from the optical redshift.

\begin{figure}
	\includegraphics[width=\columnwidth]{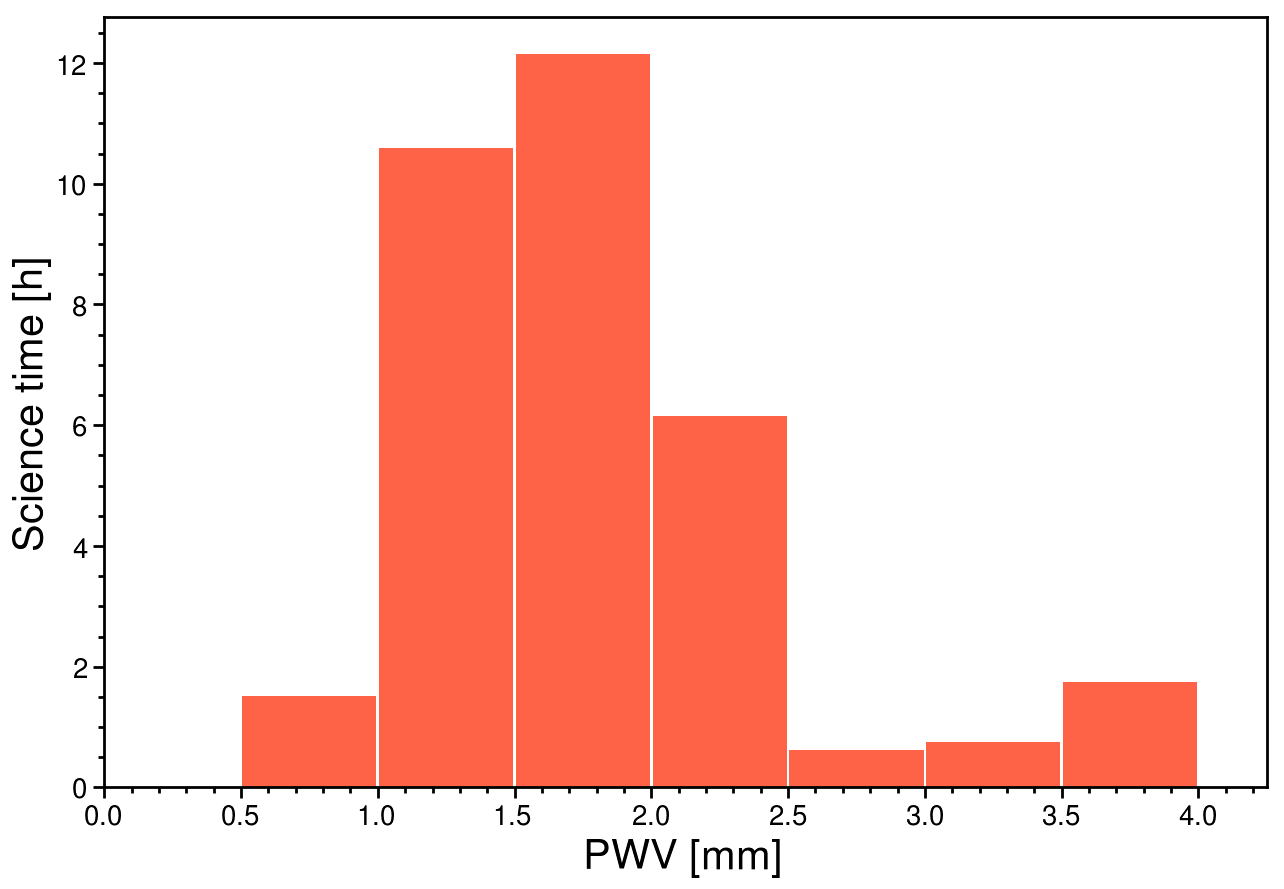}
    \caption{Distribution of the precipitable water vapour (PWV) during the science observations. The bulk of the observations were taken in excellent weather conditions below 2.5 mm PWV.}
    \label{fig:pwv_distribution}
\end{figure}

Including overheads and calibration, the programme was completed in 48\,h, spread across different days from 2018 Apr 23 to 2018 Jun 26. A total of about 33.5\,h were spent on science exposures of the 72 targets. The precipitable water vapour (PWV) varied between 0.5 and 4\,mm, with a distribution as shown in Figure~\ref{fig:pwv_distribution}.

\subsection{Data Reduction}
We reduced the data from the APEX telescope using the \code{GILDAS} software package \code{CLASS}\footnote{\url{http://www.iram.fr/IRAMFR/GILDAS}}. For each source, the PI230 instrument produced a set of 4 GHz-wide subscans covering four different spectral ranges ($221.825-225.825$ GHz, $225.625-229.625$ GHz, $237.625-241.625$ GHz, $241.425-245.425$ GHz). We first separated the subscans into their corresponding sources. For the subscans containing the expected position of the CO(2-1) emission line, we flagged the edges of the subscan and masked the region $v \in (-500, 500)$ km\,s$^{-1}$ (i.e.\ 500 km\,s$^{-1}$ either side of the expected central velocity of the CO(2-1) emission line). We then performed a baseline subtraction on each subscan. We tested fitting zero, first and second order polynomial functions, finding that a linear baseline subtraction was the most appropriate for our data. Note that each subscan has a total velocity width of $\sim5000$ km\,s$^{-1}$ and, even after masking the region with expected CO(2-1) signal, there are plenty of channels covering a wide velocity range to ensure a reliable baseline subtraction. 

Apart from the subscan of interest which contains the expected CO(2-1) signal, the other 3 subscans are not expected to cover any strong lines\footnote{According to the Splatalogue online database, \url{https://www.cv.nrao.edu/php/splat/}}. For these subscans covering different spectral ranges, we simply flagged the edges and performed the linear baseline subtraction. 

We then averaged all subscans covering the same spectral range to create four high signal-to-noise, maximum resolution spectra covering each of the four spectral ranges produced by the PI230 instrument. These averaged spectra were checked by eye before being stitched together to produce the final `signal' ($221.825-229.625$ GHz) and `image' ($237.625-245.425$ GHz) spectra. 

In order to create useful 1D spectra, the antenna temperature (corrected for atmospheric loss), $T^{\star}_{A}$, was converted into a flux density using the antenna gain factor $S_{\nu}/T^{\star}_{A}=42$ Jy K$^{-1}$ \footnote{See \url{http://www.apex-telescope.org/telescope/efficiency/index.php?yearBy=2018}} . 

We aim to compare our cluster results with studies of galaxies in the field. As discussed earlier, one excellent comparison sample is the ALLSMOG survey, which measures the CO(2-1) content in typical, main-sequence galaxies at a similar redshift to Antlia \citep[$0.01<z<0.03$,][]{2017A&A...604A..53C}. The similarities between our sample and ALLSMOG remove a number of potential variables from the comparison (such a redshift evolution, CO ladder evolution/conversions), rendering environment as the main driver for potential differences in the samples.

\begin{figure}
	\includegraphics[width=\columnwidth]{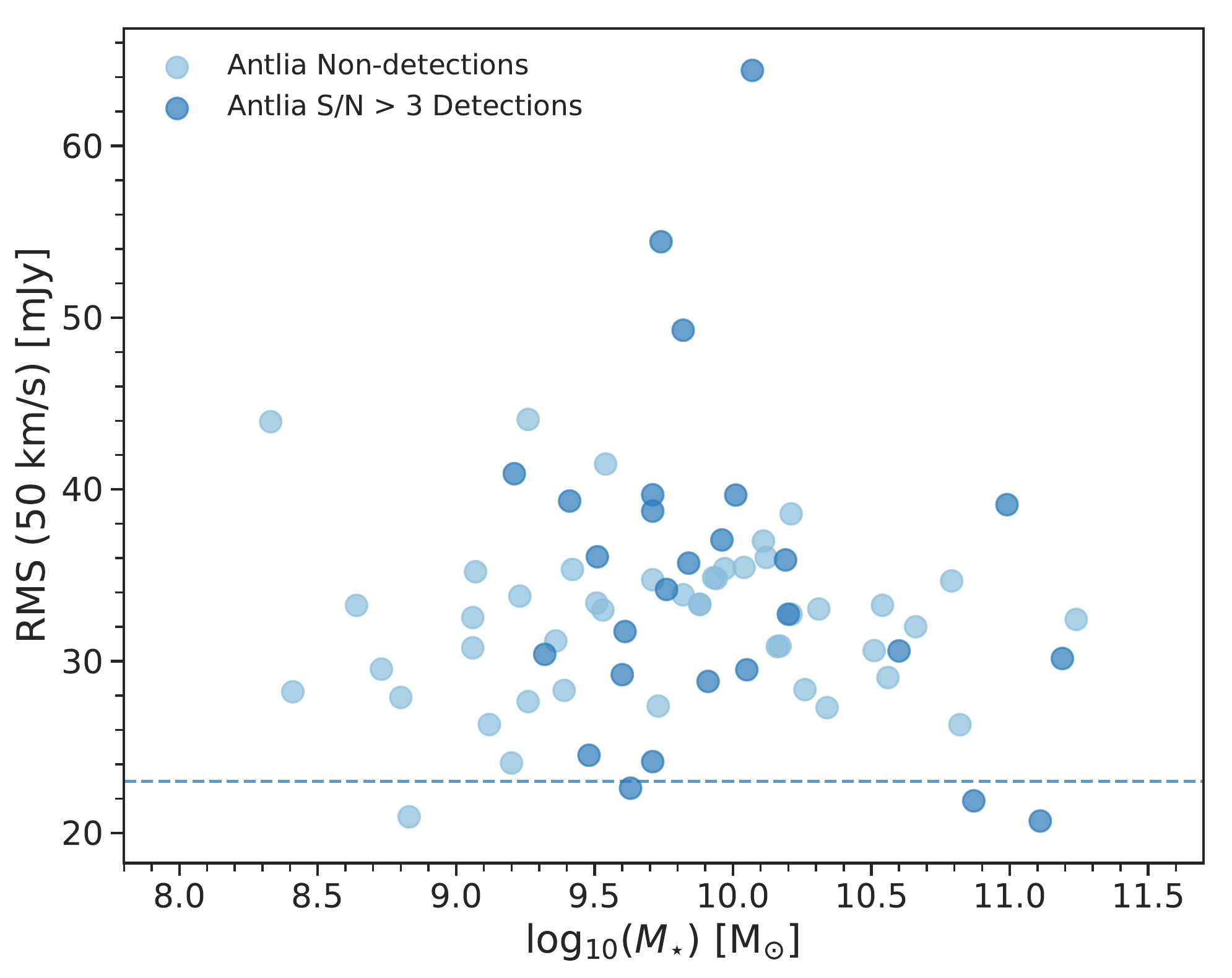}
    \caption{The 1$\sigma$ RMS noise of our 72 APEX observations. The dark blue circles represent our $S/N>3$ detections and the light blue circles represent our non-detections. The dashed line represents the survey RMS goal of $\sim$23 mJy ($\sim$0.6 mK) for spectra smoothed to a velocity resolution of 50 km s$^{-1}$.}
    \label{fig:rms_noise}
\end{figure}

Another potential comparison sample comes from the COLDGASS and xCOLDGASS surveys \citep{2011MNRAS.415...32S, 2017ApJS..233...22S}. These samples cover massive star-forming and passive galaxies with $M_{\star}>10^{10}$\,$M_{\odot}$ at $0.025<z<0.05$ with CO(1-0) observations with the 30-m IRAM telescope. Therefore, COLDGASS sample has a number of key differences to ours: resolution effects, covering different parts of the galaxies, redshift, CO transition, inclusion of AGN. \cite{2017A&A...604A..53C} compared the ALLSMOG survey to the COLDGASS and found that, given the large intrinsic scatter for star-forming galaxies, many of the trends observed by COLDGASS at higher masses are fully recovered by ALLSMOG. To minimise the number of unknowns and free parameters (e.g.\ small potential redshift evolution, uncertainty in CO conversion between CO(2-1) and CO(1-0)) and for ease of visualisation, we choose to show in our plots the ALLSMOG data points only. Relationships we plot between SFR, stellar mass and CO luminosity are derived by \cite{2017A&A...604A..53C} by using the ALLSMOG and COLDGASS data together. 

We smoothed our spectra to a velocity resolution of 50 km\,s$^{-1}$ in order to aid comparison between the RMS noise of our observations and those of ALLSMOG \citep{2017A&A...604A..53C}. Our target RMS noise of $23$\,mJy is comparable to that obtained in the APEX data from \cite{2017A&A...604A..53C} ($\sim26$\,mJy). We show in Figure \ref{fig:rms_noise} the RMS noise of our spectra as a function of stellar mass, including the survey RMS goal. We further produced spectra smoothed to different velocity resolutions in order to test whether the fits to the data were significantly affected by different resolutions. This is discussed further in Section \ref{sec:Measuring_CO_Line_Luminosity}.

The APEX CO(2-1) spectra next to optical images of the host galaxies can be found in Appendix~\ref{fig:opticalsources}.

\begin{figure*}
    \begin{center}
	\includegraphics[width=0.9\textwidth]{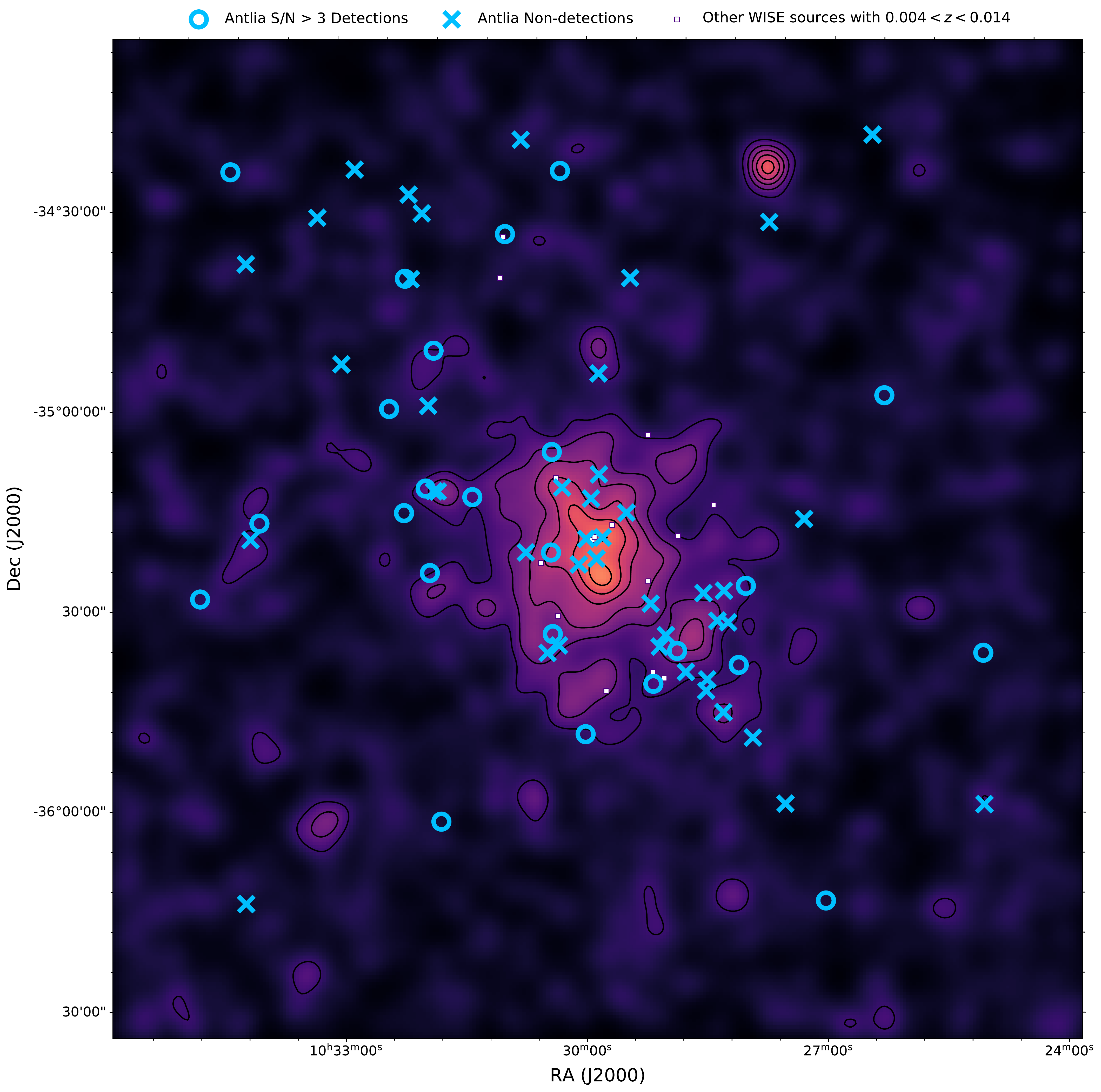}
	\end{center}
    \caption{Distribution of galaxies in the Antlia cluster that we followed up with APEX CO(2-1) observations. The background shows a smoothed ROSAT $0.5-2.4$\,keV X-ray image of the cluster. We use the symbol O to denote CO detections ($S/N \gtrsim 3$), and X to denote non-detections ($S/N < 3$). We also show in white all the sources with redshifts consistent with being associated with the cluster ($0.004<z<0.014$). The other {\it WISE}-detected cluster sources were not selected for follow-up because they did not have any significant star formation indicated by their 12\,$\mu$m emission from the \textit{WISE} data.}
    \label{fig:RA_DEC}
\end{figure*}

\section{Data Analysis}
\label{sec:Analysis}

\subsection{Fitting Line Profiles}
The first step in probing the molecular gas reservoirs of cluster members is determining their CO(2-1) line luminosity. 
For each source in our sample, we fitted both a Gaussian and a Lorentzian profile to their CO(2-1) emission line, using both \code{CLASS} and \code{LMFIT} \citep{newville_matthew_2014_11813}. For any sources that, on inspection, had CO(2-1) profiles that clearly deviated from a single peak, we also experimented with fitting both double-horned and double-Gaussian profiles. For each of these four profiles, we further experimented with fixing the continuum value to 0\,Jy (which is physically motivated as we performed a linear baseline subtraction during the data reduction process, outlined in Section \ref{sec:Data_reduction}) and with leaving the continuum value as a free parameter. For the double-Gaussian profile, we tested fitting a single standard deviation ($\sigma$) to both peaks, and fitting individual $\sigma$ values to each peak. Finally, we applied the fits to spectra smoothed to three different velocity resolutions (10\,km\,s$^{-1}$, 15\,km\,s$^{-1}$ and 20\,km\,s$^{-1}$) in order to test whether the resolution of the spectra had a significant effect on the result of the fits.

Firstly, we found that when fitting Gaussian and double-Gaussian profiles, both \code{CLASS} and \code{LMFIT} produced very similar results, irrespective of the resolution of the spectra. However, the limited functionality of \code{CLASS} means that it is not currently possible to fit a Lorentzian profile to the data, and the double horned fit proved to be extremely sensitive to the initial guesses for the parameters. For these reasons, the final fits were performed using \code{LMFIT}. We found that, for the three sets of spectra with different velocity resolutions, the fits to the data were in good agreement within the uncertainties, and we selected the spectra smoothed to a resolution of 20\,km\,s$^{-1}$ for the final fits. We also found that the fits produced by fixing the continuum value to 0\,Jy and leaving it as a free parameter were in good agreement within the uncertainties. In the final fits, we therefore fixed the continuum value to 0\,Jy.

\subsection{Defining Detections}
\label{sec:defining_detections}
\begin{figure}
    \begin{center}
	\includegraphics[width=0.49\textwidth]{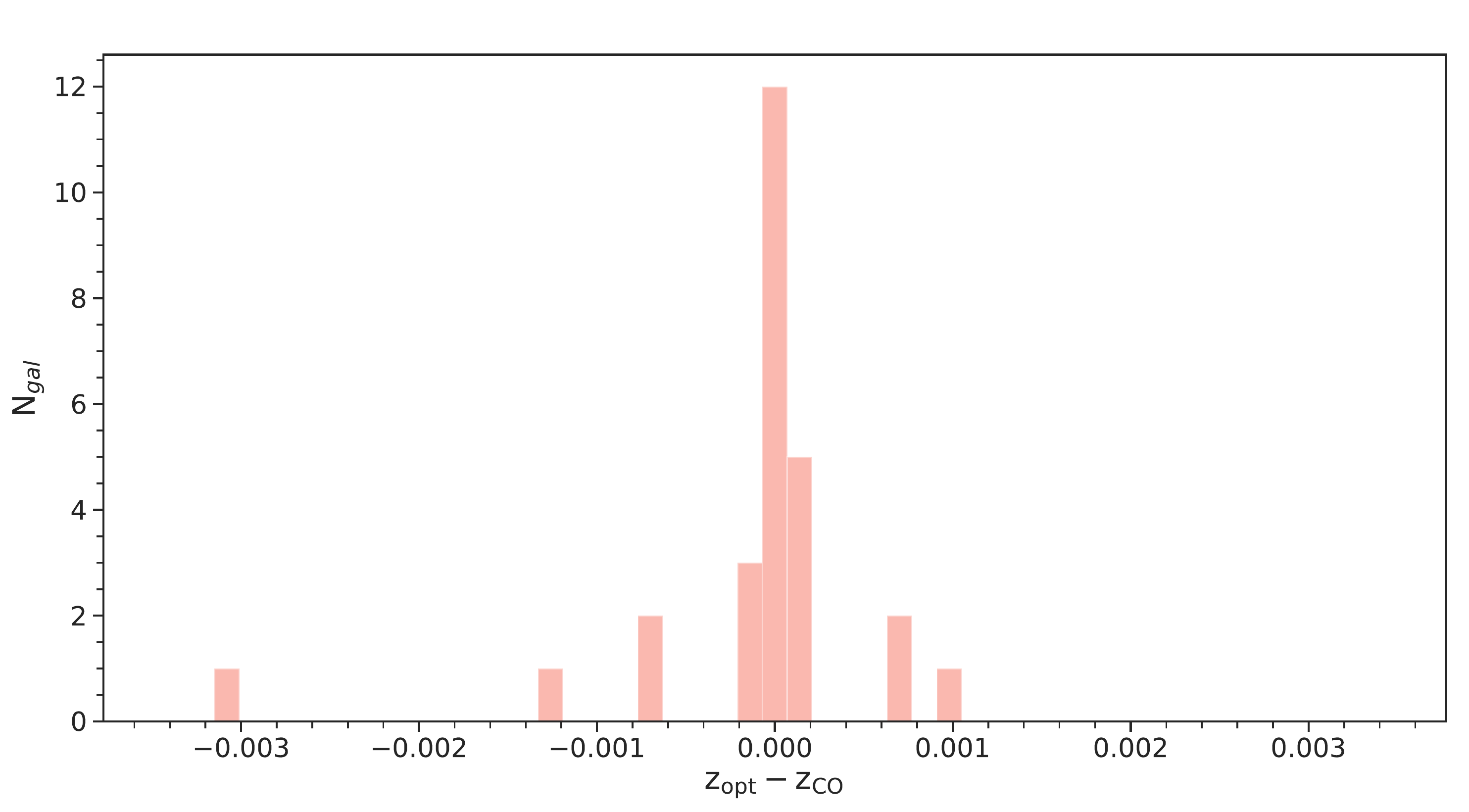}
	\end{center}
    \caption{Offset between the CO(2-1)-derived redshifts and the optical redshift for galaxies with molecular gas detections. For most galaxies, the two are in excellent agreement. The few exceptions could be caused by stripping of the molecular gas or small uncertainties in the derivation of the optical and or CO redshift.}
    \label{fig:zoffset}
\end{figure}

Of the 72 spectra in our sample, 11 showed evidence of CO(2-1) profiles inconsistent with a single peak. A double-Gaussian profile was fit to 10 of these sources, while a double-horned profile was fit to 1 source. A single Gaussian was fit to the remaining sources, including one source in which the Gaussian profile was significantly offset from the optical redshift. A distribution of the offsets between the optical and the CO redshifts is shown in Figure~\ref{fig:zoffset}.

To select our detections, we calculate the $S/N$ ratio by dividing the velocity-integrated flux density by its error estimated from the fitting procedure, and impose a $S/N > 3$ cutoff. This differs somewhat from the detection criteria for the ALLSMOG survey. \cite{2017A&A...604A..53C} similarly impose a threshold of $S/N > 3$ for their detections, but calculate the signal-to-noise ratio by dividing the peak flux of the emission line by the RMS noise of the spectra. This difference is necessary as, while \cite{2017A&A...604A..53C} fit only single Gaussians to their emission lines, we fit various other functions which makes the definition of a peak flux difficult. We note that using the velocity-integrated flux density rather than the peak flux likely results in our $S/N$ ratios being underestimated relative to those of the ALLSMOG survey.

We find CO(2-1) $S/N > 3$ detections in 27 of the 72 spectra. Our overall detection rate of $\sim37.5\%$ is lower than the ALLSMOG survey. This difference is primarily due to the stringent selection criteria for main-sequence galaxies imposed in \cite{2017A&A...604A..53C}. Additionally, we impose a strict $S/N$ cutoff for our detections, while \cite{2017A&A...604A..53C} include a number of marginal detections for which the central velocity of the CO detection agrees with the expected central velocity measurement based on the source's H{\sc i} 21cm spectrum. We note that our detection rate for sources within 1 dex of the main sequence is $86\%$, much higher than the $\sim 47\%$ achieved by the ALLSMOG survey.

\subsection{Measuring CO Line Luminosity}
\label{sec:Measuring_CO_Line_Luminosity}
For each of our detections, we take the best fit parameters from \code{LMFIT} and calculate the CO(2-1) line luminosity using the following equation:
\begin{equation}
L'_\textsc{CO(2-1)}=3.25\times10^{7} \frac{D_{L}^{2}}{\nu_{obs}^{2}\left(1+z\right)^{3}} \int S_\textsc{CO(2-1)}dv
\label{eqn:CO_luminosity}
\end{equation}
where $L'_\textsc{CO(2-1)}$ is defined as the CO(2-1) brightness temperature luminosity in units of K\,km\,s$^{-1}$\,pc$^{2}$, $D_{L}^{2}$ is the luminosity distance measured in Mpc, $\nu_{obs}$ is the observed central frequency of the CO(2-1) emission line in GHz, $z$ is the cluster redshift and $\int S_\textsc{CO(2-1)}dv$ is the velocity-integrated flux density measured in Jy\,km\,s$^{-1}$ \citep{1997ApJ...478..144S}. This velocity-integrated flux density is corrected for the possible loss of flux due to CO emission falling outside of the APEX beam. \cite{2017A&A...604A..53C} find that this correction is typically very small, and so we simply divided the velocity-integrated flux density of each source by their median correction factor of 0.98. For our non-detections, we again followed \cite{2017A&A...604A..53C} and estimated informative upper limits on the velocity-integrated flux density using the following equation:
\begin{equation}
\int S_\textsc{CO(2-1)}dv=3\sigma_\textsc{RMS}\sqrt{\delta v \Delta v_\textsc{CO(2-1)}}
\label{eqn: upper_lims}
\end{equation}
where $\sigma_\textsc{RMS}$ is the RMS noise of the spectrum smoothed to a velocity resolution of $\delta v$ and $\Delta v_\textsc{CO(2-1)}$ is the expected CO(2-1) line width, which we defined as the average FWHM of our CO(2-1) detections ($\Delta v_\textsc{CO(2-1)} \sim 170$ km s$^{-1}$). This upper limit on the velocity-integrated flux density was then applied to Equation \ref{eqn:CO_luminosity} to estimate upper limits on the CO(2-1) luminosity.

Typically, the luminosity of the CO(1-0) transition is used to estimate molecular gas mass. We therefore converted our CO(2-1) luminosities into CO(1-0) luminosities using the following relation:
\begin{equation}
r_{21} = \frac{L'_\textsc{CO(2-1)}}{L'_\textsc{CO(1-0)}}
\label{eqn: L_ratio}
\end{equation}
where we assume a value of $r_{21}=0.8 \pm 0.2$ based on observations of local star-forming spirals \citep{1993A&AS...97..887B,2009AJ....137.4670L} and $z \sim 1.5-2.0$ star-forming disks \citep{2010ApJ...718..177A,2014MNRAS.442..558A}. 

In order to convert the CO(1-0) luminosities into estimates for the molecular gas mass, we use the equation:
\begin{equation}
    M_{\text{mol}} = \alpha_\textsc{CO} L_\textsc{CO(1-0)}
\end{equation}
where $\alpha_\textsc{CO}$ is the CO-to-H$_{2}$ conversion factor. Determining the value of $\alpha_\textsc{CO}$ is still an active area of research, and while a constant value of $\alpha_\textsc{CO}$ is often applied based on observations of nearby galaxies, $\alpha_\textsc{CO}$ is likely to be strongly dependent on the conditions of the local ISM, including pressure, gas dynamics and metallicity \citep{2013ARA&A..51..105C}. Theoretical models \citep[e.g.][]{2010ApJ...716.1191W,2011MNRAS.412..337G} further predict that $\alpha_\textsc{CO}$ is a strong function of dust extinction $A_{V}$. In order to accurately estimate $\alpha_\textsc{CO}$ for our sample, it is therefore necessary to carry out further observations that aim to characterise the spatial distribution of the molecular gas, as well as the metallicity and dust content of the local ISM. In order to better compare with the sample of field galaxies, we follow \cite{2017A&A...604A..53C} and select a Milky-Way type CO-to-H$_{2}$ factor of $\alpha_\textsc{CO} = 4.3$\,M$_{\odot}$ \citep{2013ARA&A..51..207B}, but note that we typically consider the CO(1-0) luminosity as a proxy for the molecular gas mass throughout this paper. 

\begin{figure}
	\includegraphics[width=\columnwidth]{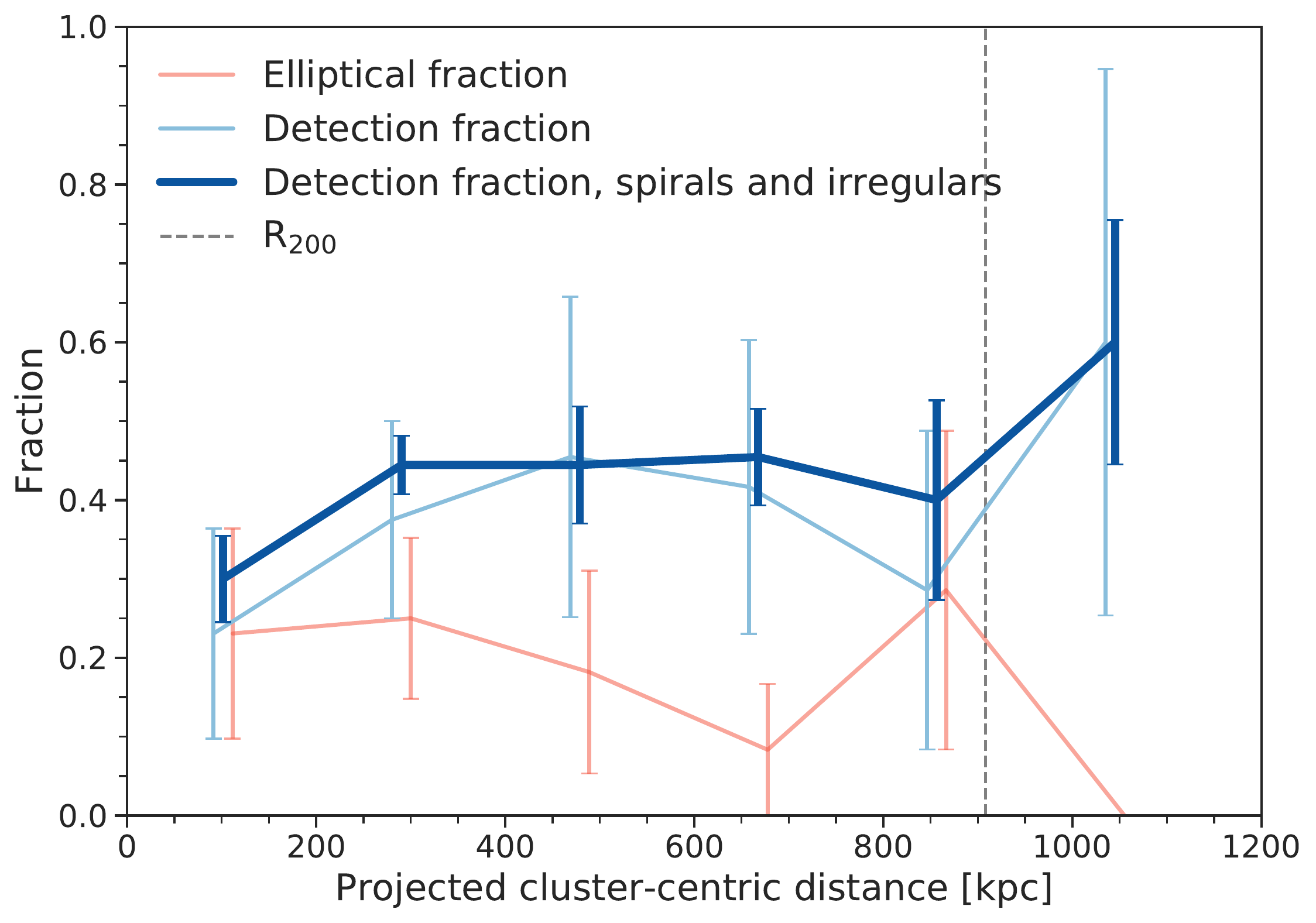}
    \caption{Detection rate as function of projected cluster-centric radius. On average the detection rate increases with distance from the cluster core. This is mostly driven by the fraction of ellipticals in our sample of targeted sources. However, the detection rate for sources visually classified as spirals and irregulars also slightly increases with cluster-centric distance.}
    \label{fig:projradius}
\end{figure}

\section{Molecular Gas Detections in the Antlia Cluster}
\label{sec:gas_detections}

As discussed, we followed up a total of 72 sources with APEX molecular gas observations. Figure~\ref{fig:RA_DEC} shows a spatial distribution of sources with redshifts confirmed to be within the $0.004-0.014$ range, together with all the sources followed up as part of our optical selection.

\begin{figure*}
\centering
\subfloat{%
  \includegraphics[width=0.4\textwidth]{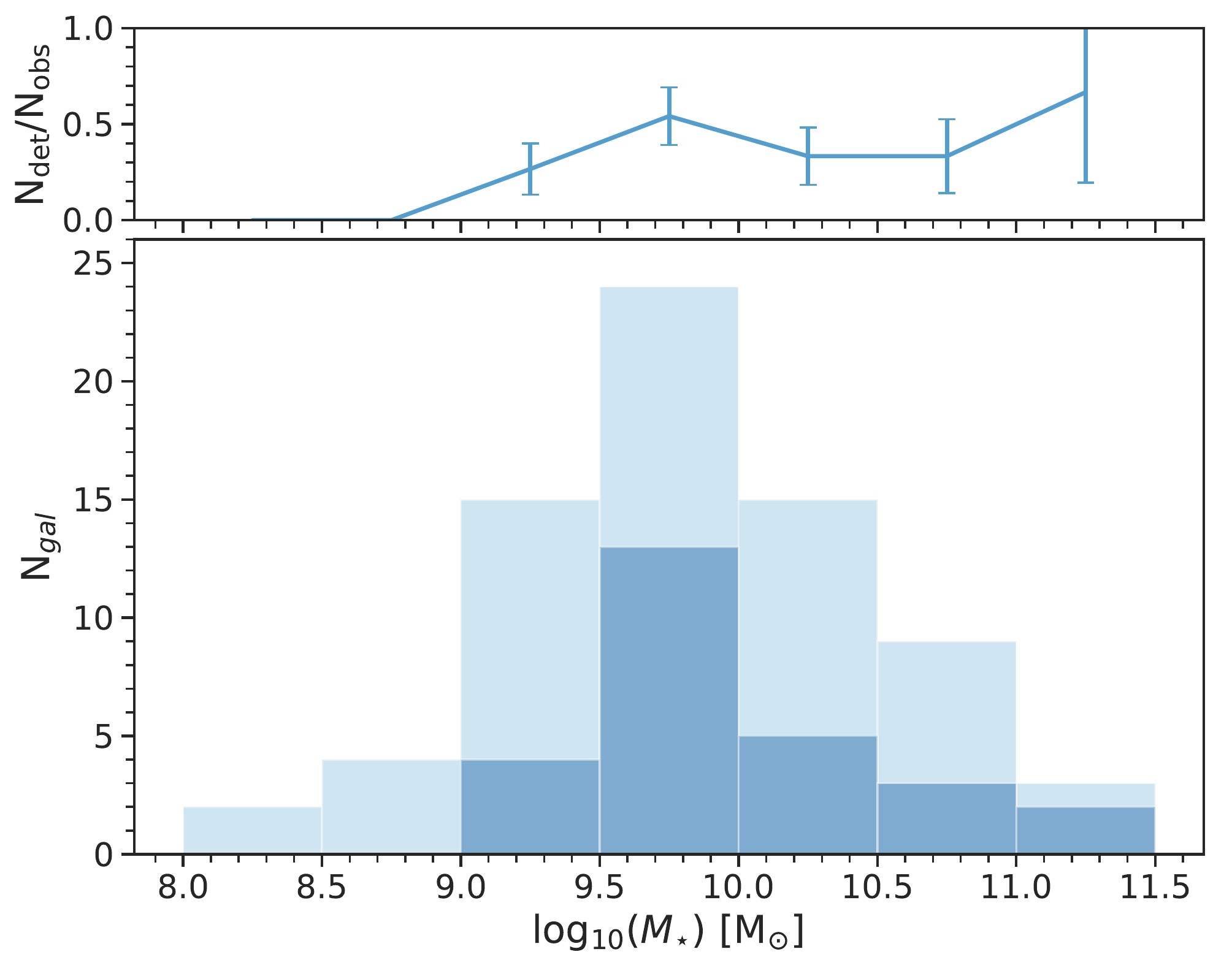} \hspace{0.4cm} 
}%
\subfloat{%
  \includegraphics[width=0.4\textwidth]{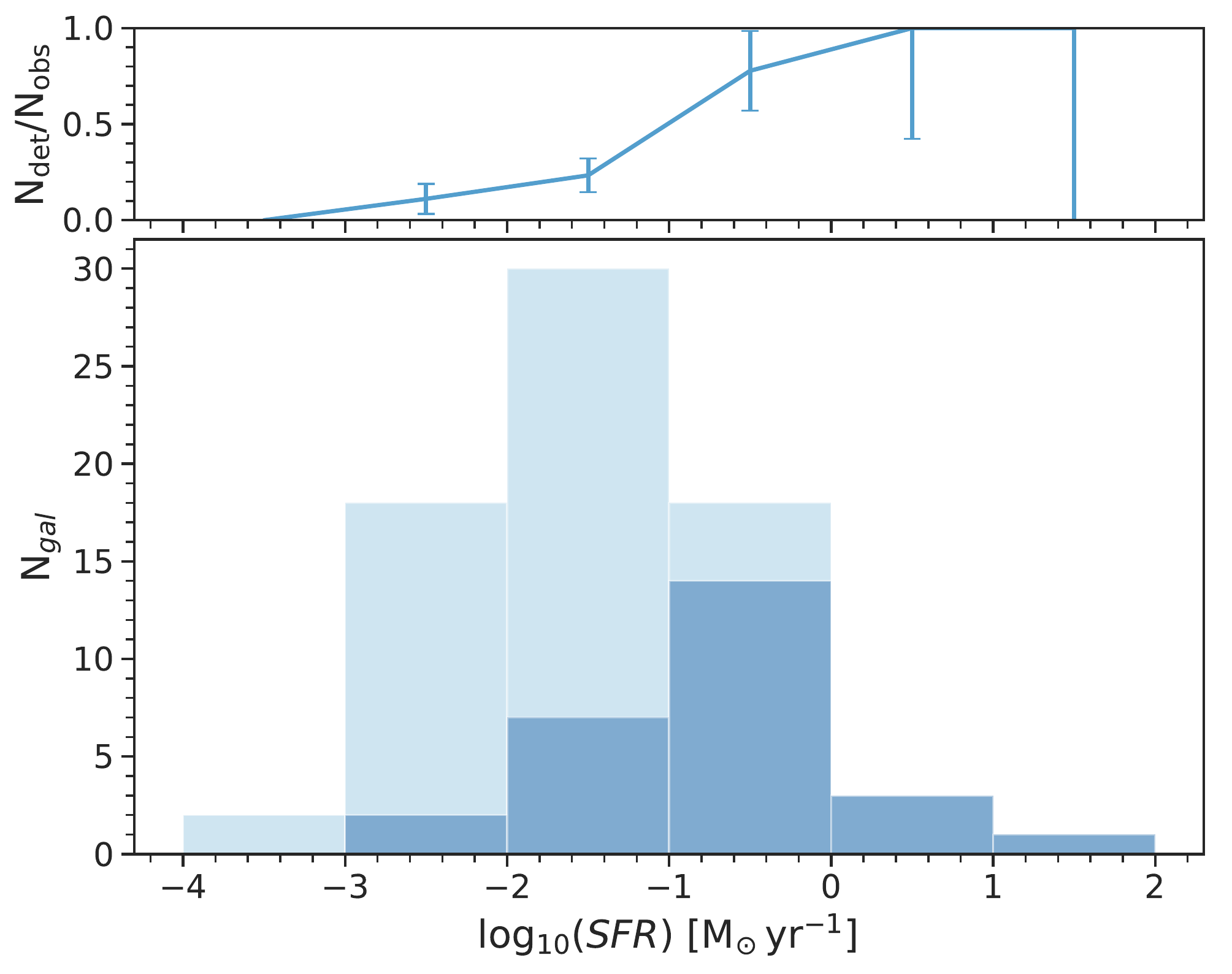}%
}
\caption{The distribution of detections and non-detections from our APEX observations as a function of stellar mass (\textit{left}) and SFR (\textit{right}). In the lower panels of these plots, the darker bars represent the number of detections, while the lighter bars represent the number of non-detections. The upper panels show the detection fraction in different bins with their associated Poisson errors. The detection fraction increases with both mass and SFR, though we note that the correlation between detection fraction and SFR is stronger than the correlation between detection fraction and stellar mass.}
\label{fig:detection_rates}
\end{figure*}

Out of the 72 optically-selected sources, we obtain 27 detections, mostly clustered towards the high star formation and high stellar mass end (see discussion in Section~\ref{sec:Detection_Rates}). In this section, we present both the CO(2-1) detection rates and CO(1-0) luminosities of our sample as a function of cluster and galaxy properties.

\subsection{Detection Rates as a Function of Radial Distance and Galaxy Type}
By focusing on the optically selected sources, we investigated the detection rate as function of projected cluster-centric distance (Figure~\ref{fig:projradius}). On average, the detection rate increases with projected cluster-centric distance, from about 0.3 to 0.6. This is driven by the preferential location of galaxies with lower star formation within 500 kpc of the cluster core. These low-SFR galaxies have also been optically classified as elliptical based on their HST/VIMOS/DSS morphology. The detection fraction of spirals and irregulars drops at the very core within $\sim200$\,kpc and rises a bit beyond the $R_{200}$ of the cluster, but otherwise is constant between $200-900$\,kpc. Out of the 27 detections, 25 are hosted in galaxies optically classified as spirals or irregulars, while only 2 are in galaxies classified as ellipticals.

We attempted to quantify the significance of this result. As a first test, we fitted a constant function to the four data points comprising the $200-900$\,kpc region where the detection fraction remains roughly constant using simple \code{SciPy}\footnote{\url{https://www.scipy.org/}} packages, and calculated the distance (in $\sigma$) between the first data point and this line. We found that the first data point resides $\sim1.1\sigma$ below this constant relation, and so is only marginally inconsistent with a flat relation between detection fraction and cluster-centric radius. As a slightly more robust test, we fitted both linear and constant models to all six data points using \code{LMFIT}, comparing the chi-squared statistic ($\chi^{2}$) for each fit. For the constant model, the best fit value was $0.331 \pm 0.036$, while the linear model favoured a gradient of $(2.03 \pm 1.42) \times 10^{-4}$\,kpc$^{-1}$. We determine $\chi^{2}$ values of 1.25 and 1.93 for the linear and constant fits respectively, indicating that the data slightly favour an increasing detection fraction with increasing cluster-centric radius. We also test fitting to different numbers of bins, finding that a positive linear correlation is always favoured over a constant fit. This result is further consistent with previous works which show that galaxies residing in the outskirts of massive clusters tend to be more blue and star-forming. For example, \cite{2019A&A...622A.117S} find that, as early as $z\sim1.5$, the most massive clusters effectively suppress star formation in their central regions. Similarly, \cite{2014A&A...571A..80A} find that, in the $z=0.44$ cluster M1206, tidal disruption near the cluster core can convert star-forming galaxies into passive ones, and can even destroy galaxies entirely.

\subsection{Detection Rates as a Function of Galaxy Properties}
\label{sec:Detection_Rates}

The distribution of detections as a function of source redshift is shown in Figure~\ref{fig:redshift_histogram}. Given the low number statistics, we can only study the CO(2-1) detection rate in rather large redshift bins. The fraction of CO(2-1) detections remains reasonably constant throughout the cluster. We present the distribution of our detections as a function of stellar mass and SFR in Figure \ref{fig:detection_rates}. In the lower panel of these plots, the darker sections of the bars represent the number of detections in each bin, while the lighter regions represent the non-detections. The upper panels show the fraction of our CO(2-1) detections in each bin (i.e.\ the number of detections divided by the total number of observations in each bin) as a function of the galaxy parameter, with the associated Poisson error. From Figure \ref{fig:detection_rates}, we infer that there is a clear evolution in the CO(2-1) detection rate with both stellar mass and SFR. There are no CO(2-1) detections in our sources with stellar masses less than $\sim10^{9}$\,M$_{\odot}$. At stellar masses above $\sim10^{10}$\,M$_{\odot}$, our low detection rate is driven by the large fraction of passive, elliptical galaxies in our sample. The detection rate for purely star-forming galaxies, selected within 1\,dex of the main sequence, is 86\%. By contrast, for a similar sample of star-forming galaxies in the field, \cite{2017A&A...604A..53C} find that their detections cluster towards more massive galaxies ($> 10^{9.5}$\,M$_{\odot}$). Similarly, the figure shows a smooth increase in the CO(2-1) detection fraction with increasing SFR. This increase is expected, as Figure \ref{fig:smass_sfr} demonstrates that stellar mass and SFR are correlated in our sample, although the correlation between SFR and detection fraction is noticably stronger than the correlation between stellar mass and detection fraction. We infer that the majority of our detections are clustered towards the more massive and more star forming galaxies in our sample, while a significant fraction of our sample have quite low SFRs, as demonstrated in Figure~\ref{fig:smass_sfr}.

\subsection{CO Luminosity as a Function of Galaxy Properties}
\label{sec:CO_Luminosity_Correlations}

\begin{figure*}
\begin{minipage}{.75\linewidth}
\centering
\subfloat[]{\label{main:a}\includegraphics[width=0.99\columnwidth]{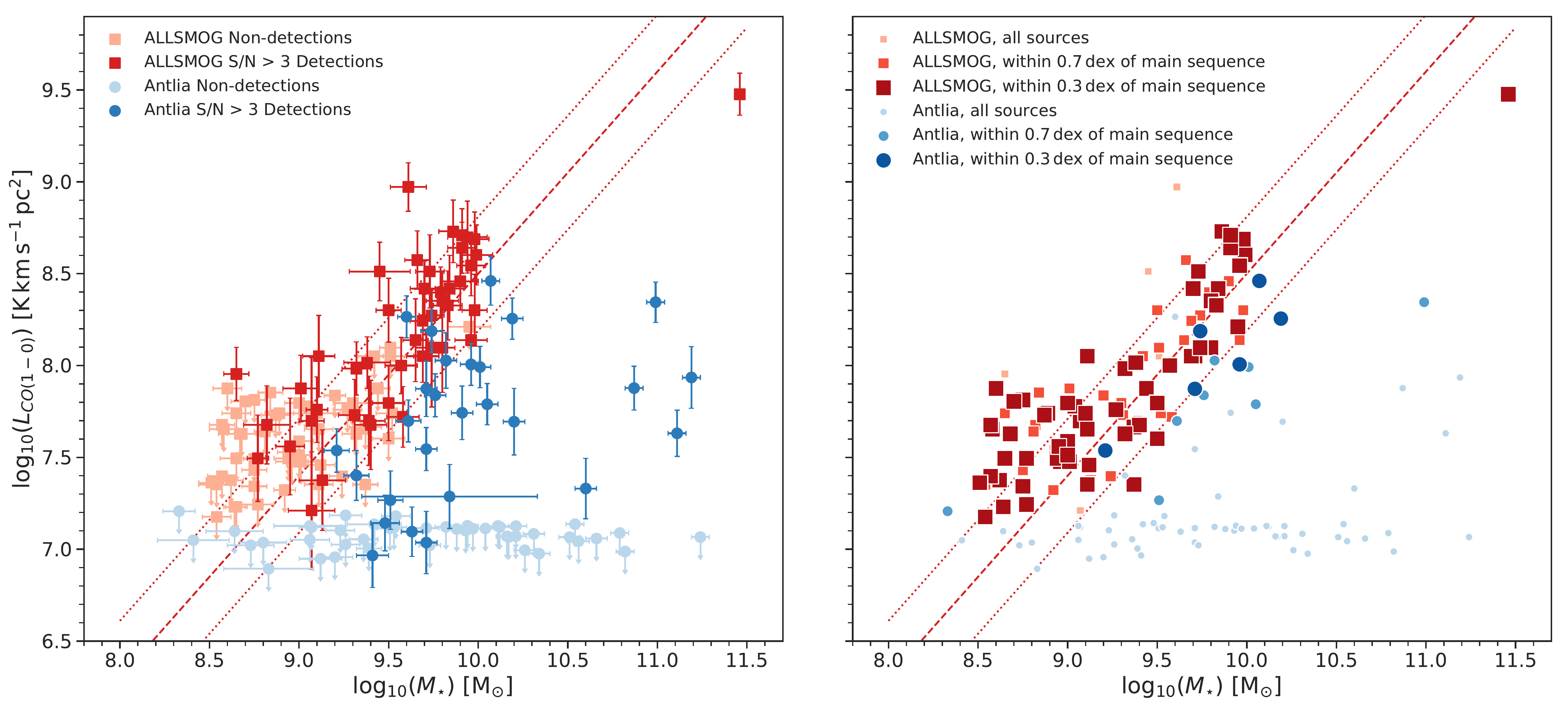}}\\
\subfloat[]{\label{main:b}\includegraphics[width=0.99\columnwidth]{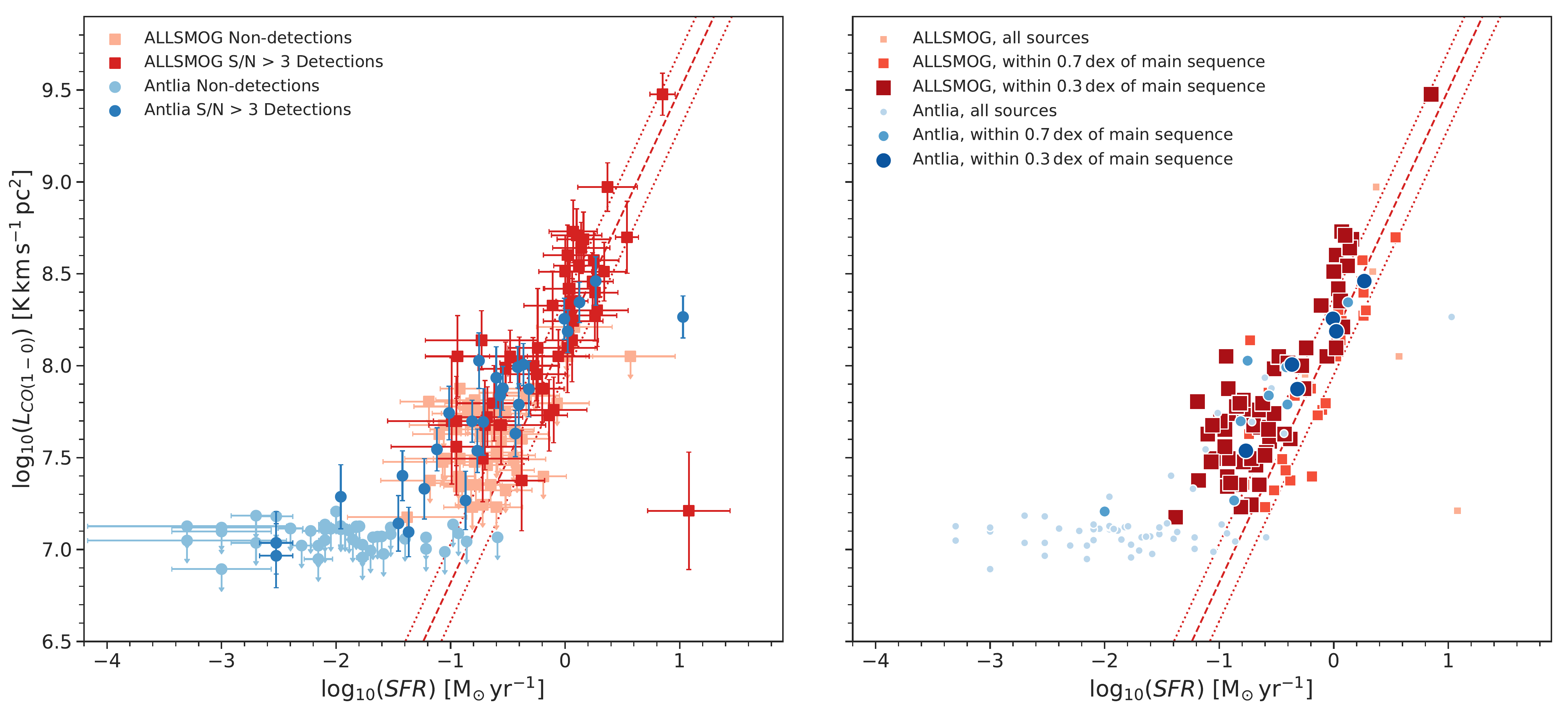}}
\end{minipage}%
\centering
\caption{The CO(1-0) line luminosity as a function of stellar mass (\textit{upper panel}) and SFR (\textit{lower panel}). In the left panels, darker blue points represent the $S/N>2.5$ detections, while the lighter blue points represent the upper limits for non-detections. The darker red squares represent the CO(1-0) luminosities for the $S/N>3$ sources in the ALLSMOG survey of galaxies in the field, while the lighter red squares represent their non-detections. The dashed red line shows the best fit relation for the ALLSMOG survey, while the red dotted lines show the intrinsic scatter to their data \citep{2017A&A...604A..53C}. The right hand plots encode the location of the detections and upper-limits with respect to the main sequence. Main sequence cluster galaxies (selected within 0.3\,dex or 0.7\,dex of the main sequence) fall on the relation derived for their field counterparts. Galaxies with lower SFRs, located below the main sequence, fall to the right of the relation, i.e.\ they have less molecular gas compared to a galaxy of similar mass with higher SFR located on the main sequence. This illustrates the three dimensional relationship between molecular gas, SFR and stellar mass.}
\label{fig:CO_luminosity_plots}
\end{figure*}

In this section, we investigate the relationship between CO(1-0) line luminosity and galaxy properties (i.e.\ SFR and stellar mass). In order to highlight the effects that disturbed cluster environments have on the molecular gas content of their member galaxies, these results will be compared to those of \cite{2017A&A...604A..53C}, who undertake a similar analysis on their sample of field galaxies.

It is important to note that a direct comparison between our results and those from the ALLSMOG survey is made difficult by the difference in selection criteria between the two studies. \cite{2017A&A...604A..53C} select only those galaxies with stellar masses in the range  $10^{8.5}$\,M$_{\odot}<M_{\star}<10^{10}$\,M$_\odot$, while we require our galaxies to have stellar masses $M_{\star}>10^8$\,M$_\odot$, and we impose no upper limit. Moreover, \cite{2017A&A...604A..53C} select only those galaxies defined as star forming based on their position on the $\log($[{\sc Oiii}]$/$H$\beta)$ vs. $\log($[{\sc Nii}]$/$H$\alpha)$ diagram, while our cut of SFR$>0.0005$ M$_{\odot}$ yr$^{-1}$ is based on 12$\mu$m emission from the \textit{WISE} W3 band. Effectively, \cite{2017A&A...604A..53C} select star-forming galaxies within $\sim1$\,dex of the main sequence, while our selection is broader and it includes galaxies on and below the main sequence. \cite{2017A&A...604A..53C} make a further cut based on metallicity, while we currently have no metallicity measurements for our sample of galaxies. We also note that \cite{2017A&A...604A..53C} impose a slightly different signal-to-noise definition, as discussed in Section \ref{sec:Measuring_CO_Line_Luminosity}. 

In Figure~\ref{fig:CO_luminosity_plots}, we present the relation between CO(1-0) luminosity and stellar mass. \cite{2017A&A...604A..53C} report strong $L_{\text{CO}(1-0)}-M_{\star}$ and $L_{\text{CO}(1-0)}-$SFR relations for main-sequence galaxies in the field. Considering all of our detections, we could infer that there is no convincing trend between CO(1-0) luminosity and stellar mass or between CO(1-0) luminosity and SFR for galaxies in the Antlia cluster. The relations from \cite{2017A&A...604A..53C} thus seem to hold only for star-forming galaxies, and break down when extrapolated below the main sequence. This further suggests that the molecular gas reservoirs depend on both SFR and stellar mass and that galaxies are actually located on a stellar mass-SFR-molecular gas plane.

Close inspection of the upper-right plot in Figure~\ref{fig:CO_luminosity_plots} unveils a trend between the position of our sources in the individual $L_\textsc{CO(1-0)}$ vs. $M_{\star}$ and $L_\textsc{CO(1-0)}$ vs. SFR plots, and their location relative to the main sequence in the $M_{\star}-$SFR plane. The closer to the main sequence an Antlia cluster galaxy is, the closer it moves to the field relation derived in \cite{2017A&A...604A..53C}. Thus, galaxies located within 0.7\,dex above or below the main sequence are well within the scatter of the relationship between stellar mass and CO(1-0) luminosity. We find that, for galaxies in our sample located below the main sequence, the majority lie below the scatter of the trend inferred for field galaxies \citep{2017A&A...604A..53C}. This is particularly true for higher mass sources ($\sim10^{10.5}$\,$M_{\odot}$), which are all passive ellipticals with SFRs $\sim100$ times lower than in main sequence galaxies. The five detections in more massive ($\gtrsim10^{10.5}$\,$M_{\odot}$) sources show comparable CO(1-0) luminosities to their lower mass counterparts, and hence lie further below the relation of \cite{2017A&A...604A..53C}. \cite{2017A&A...604A..53C} further discuss that their sample selection likely misses galaxies residing below the main sequence which, in turn, are expected to lie below the $L_{\text{CO}(1-0)}-M_{\star}$ trend. We can therefore conclude that it is likely that the CO(1-0) luminosities of the majority of our galaxies are consistent with those of field galaxies lying just below the main-sequence.

Interestingly, the lower panels of Figure \ref{fig:CO_luminosity_plots} demonstrate a strong $L_{\text{CO}(1-0)}-$SFR trend not only for main-sequence galaxies, but for all Antlia cluster galaxies. This is similar to the results of \cite{2017A&A...604A..53C}, who discuss that the tight linear relation between SFR and CO(1-0) luminosity is expected and likely the result of the well documented empirical relation between the surface density of cold molecular gas and SFR (the Schmidt-Kennicut law; \citealt{1959ApJ...129..243S,1998ApJ...498..541K}). We further find that the vast majority of our detections with SFRs in the range probed by the ALLSMOG survey lie within or very close to the intrinsic scatter of the $L_{\text{CO}(1-0)}-$SFR relation inferred by \cite{2017A&A...604A..53C}. We conclude that, for a given SFR, galaxies in the Antlia cluster have comparable CO(1-0) luminosities to their counterparts in the field, irrespective of the mass of the galaxy. We identify one Antlia galaxy in the lower panels of Figure \ref{fig:CO_luminosity_plots} that is clearly distinct from the rest of the population, with a SFR roughly an order of magnitude larger than the next most star-forming source. This source (J103152) has a stellar mass placing it towards the lower end of the stellar mass distribution of our sample (M$_{\star} = 10^{9.6\pm0.05}$\,M$_{\odot}$), and shows some of the strongest evidence for disturbance of any of the Antlia cluster galaxies. This source is discussed further in Section \ref{sec:Discussion:Quenching}.

\begin{figure}
	\includegraphics[width=\columnwidth]{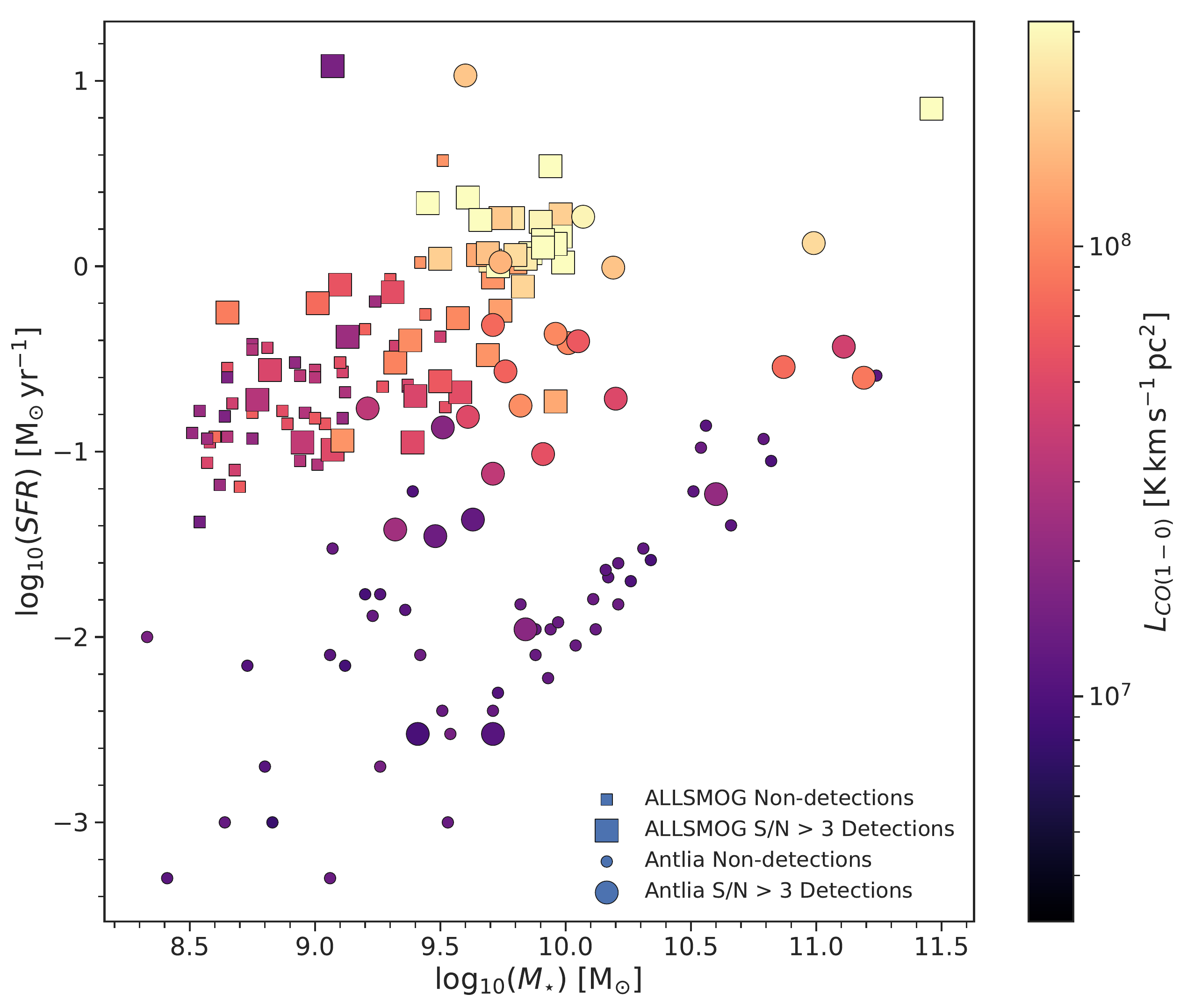}
    \caption{Distribution of molecular gas within the stellar mass SFR plane. We show the detections in the larger symbols and the non-detection upper limits in the smaller symbol. Both our survey for galaxies within the Antlia cluster and the survey of field star forming galaxies (ALLSMOG) are shown. Note than our data is deeper than ALLSMOG, thus our upper limits are more stringent. We find that cluster galaxies have a similar amount of CO(1-0) as field galaxies with a similar stellar mass and star formation rate.}
    \label{fig:co_smass_sfr}
\end{figure}

To facilitate a comparison between the molecular gas content of galaxies in the field and galaxies in a disturbed cluster environment, it is more appropriate to select galaxies in the two surveys with both similar stellar masses and SFRs, and compare the CO(1-0) luminosities of the two. To this end, we present in Figure \ref{fig:co_smass_sfr} the distribution of our sources in the $M_{\star}-{\rm SFR}-L_{\text{CO}(1-0)}$ plane alongside the sources in the ALLSMOG survey, with the color of the points indicating the CO(1-0) luminosity. We find that, while a significant fraction of our galaxies occupy a distinct parameter space in the $M_{\star}-$SFR plane, sources in the Antlia cluster and in the field with similar stellar masses and SFRs also have comparable CO(1-0) luminosities, and hence comparable reservoirs of molecular gas. The molecular gas reservoirs are proportional to both the SFR and the stellar mass, but do not significantly depend on environment. Thus, the figure reveals that galaxies in both field and cluster environments lie on a single stellar mass-SFR-molecular gas plane. This non-trivial three dimensional relation is illustrated in the projections shown previously in Figure~\ref{fig:CO_luminosity_plots}.
The implications will be discussed further in Section \ref{sec:Discussion:Molecular_Gas_Content}.
 
\subsection{Non-Gaussian Molecular Gas Profiles as a Function of Radial Distance}
\label{sec:non_gaussian_distance}

In this section, we focus on the galaxies in our sample that show evidence of disturbed molecular gas reservoirs, inferred from their non-Gaussian CO(2-1) emission line profiles. In Figure \ref{fig:non_gaussian_distance}, we plot the number of non-Gaussian CO(2-1) detections as a function of projected cluster-centric distance in order to determine where preferentially in the cluster these galaxies with disturbed molecular gas reservoirs reside. We infer that there is a possible evolution in the number of sources with non-Gaussian CO(2-1) profiles with projected cluster-centric distance. While the number of detections peaks at $\sim400$\,kpc and decreases smoothly out to $\sim1000$\,kpc, the number of our sources with non-Gaussian CO(2-1) emission profiles peaks a little further out at $\sim600$\,kpc, with just two sources displaying non-Gaussian CO(2-1) profiles out of a total of 9 detected sources at a distance of $\sim400$\,kpc. Outside of 600\,kpc we find that two sources out of a total of six detected sources have CO(2-1) emission profiles that are non-Gaussian. We discuss the physical interpretation of these results in Section \ref{sec:Discussion:Quenching}.

\section{Discussion}
\label{sec:Discussion}

\subsection{Molecular and Atomic Gas in Relaxed Clusters}

It has long been established that cluster environments can have a strong effect on the atomic gas properties of cluster galaxies. For example, cluster spirals, at fixed mass, are significantly (up to a factor of 10 times) deficient in H{\sc i} gas when compared to samples of similarly-selected field spirals. The magnitude of the effect depends on the X-ray properties of the cluster, indicating a more pronounced effect for more massive, luminous, relaxed clusters \citep[e.g.][]{1983AJ.....88..881G, 1985ApJ...292..404G, 2010AJ....139.2716C}. SFR measured from tracers of recent, massive, newly born stars does not correlate with H{\sc i}: the SFR does not seem to be directly suppressed in galaxies that have lost their H{\sc i} gas, indicating that the gas phase that directly feeds SF, the molecular gas, is not as heavily affected by the harsh cluster environment as the neutral atomic gas \citep[e.g.][]{1984AJ.....89.1279K}. The stronger depletion of H{\sc i} gas has been attributed to a broader distribution of atomic gas that is more easily stripped by the intra-cluster medium ram pressure. 

What drives the strong morphological transformation from the field population dominated by star-forming galaxies to the predominantly passive elliptical samples in relaxed clusters? The gas surface density correlates with the SFR (i.e.\ the Kennicutt-Schmidt law) and this is interpreted as evidence for a direct link between SFR and molecular gas, i.e.\ molecular gas is converted to form new stars. One would naively expect that if cluster galaxies are on average less star forming, this is directly related to lower molecular reservoirs, caused by tidal interactions with other galaxies, stripping or simply a shut-down of fresh gas accretion.

While the properties of H{\sc i} in cluster galaxies have been established, the molecular gas properties, and particularly, how the reservoirs of molecular gas are affected by the cluster environment is still debated. Early work \citep[e.g.][]{1989ApJ...344..171K, 1995A&AS..110..521B, 1997A&A...327..522B} indicated that galaxies in the Virgo and Coma cluster display minimal evidence for molecular gas deficiency, even in cases of severe H{\sc i} deficiency. The authors initially concluded that even in massive relaxed clusters, the ram pressure is not significant enough to disturb the molecular gas reservoirs of the cluster members. The molecular gas is thought to be concentrated towards the core where the gravitational potential of the galaxy dominates, and this can effectively shield the molecular gas from the effects of the larger scale cluster environment. More recently, a number of studies using larger samples and resolved observations have found that H{\sc i}-deficient spirals of a given stellar mass residing in clusters have reduced molecular gas reservoirs compared to the field \citep[e.g ][]{2006A&A...448...29B, 2009ApJ...697.1811F, 2013A&A...557A.103J, 2014A&A...564A..67B, 2016MNRAS.456.4384M, 2019MNRAS.483.2251Z, 2017MNRAS.466.1382L}. In particular, \citet{2014A&A...564A..67B} conclude that, while the molecular gas supply is cut off and the gas is consumed through SF, the main cause for the deficiency is actually removal of gas through ram pressure. However, the molecular gas is not depleted as efficiently as the atomic gas \citep{2014A&A...564A..67B}.

\begin{figure}[t]
	\includegraphics[width=\columnwidth]{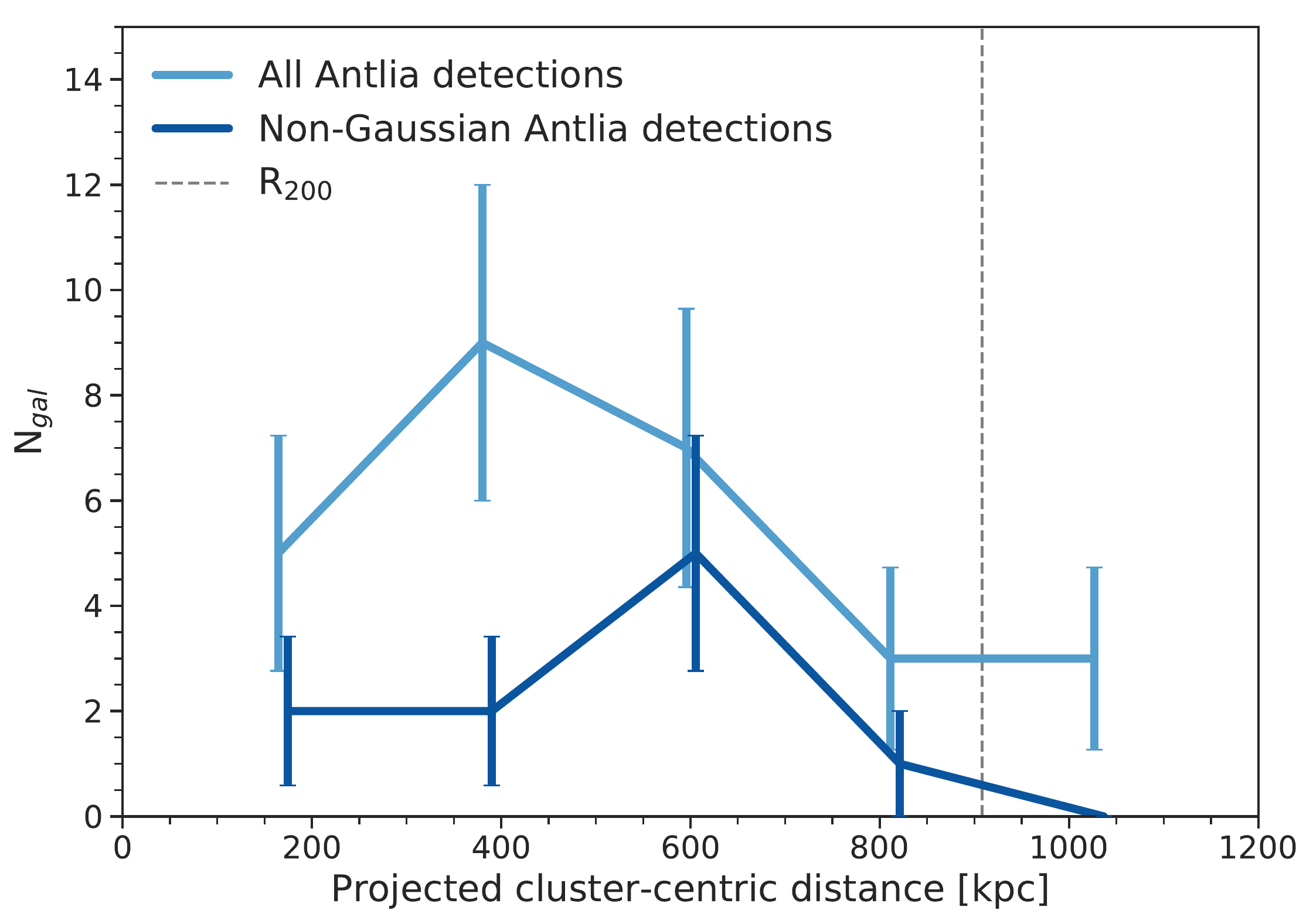}
    \caption{The number of our sources with non-Gaussian CO(2-1) profiles (dark blue), and the number of all of our detections (light blue), as a function of projected cluster-centric radius. Bearing in mind the large error bars, the number of sources with non-Gaussian CO(2-1) profiles seems to peak at $\sim600$\,kpc while the total number of detections seems to peak closer to the core of the cluster ($\sim400$\,kpc).}
    \label{fig:non_gaussian_distance}
\end{figure}

\subsection{The Molecular Gas Content of Galaxies within Disturbed Clusters}
\label{sec:Discussion:Molecular_Gas_Content}

Therefore, the molecular gas reservoirs in star forming galaxies within massive, relaxed clusters are reduced compared to their counterparts in the field. If clusters evolve from lower density environments rich in galaxies with significant gas reservoirs to massive, gas-poor clusters, should we then expect a lower (or no) molecular gas deficiency in galaxies located in disturbed clusters that are in the process of formation?

In order to evaluate the molecular gas content of galaxies in disturbed cluster environments, we compare in Figure \ref{fig:CO_luminosity_plots} the CO(1-0) luminosities of our sample as a function of stellar mass and SFR with the ALLSMOG survey of galaxies in the field \citep{2017A&A...604A..53C}. While we find a strong $L_{\text{CO}(1-0)}-$SFR relation that agrees well with galaxies in the field, we infer from our $L_{\text{CO}(1-0)}-M_{\star}$ that we are primarily probing galaxies that lie just below the main sequence (this is also evident from Figure \ref{fig:smass_sfr}), which means that a direct comparison with the ALLSMOG survey is difficult. In order to get a clearer picture of how our sample of galaxies in a disturbed cluster environment compare to those in the field, we presented in Figure \ref{fig:co_smass_sfr} a comparison between our sample and that of \cite{2017A&A...604A..53C} in the $M_{\star}-SFR-L_{\text{CO}(1-0)}$ plane, finding that galaxies of fixed stellar mass and SFR have comparable molecular gas reservoirs, regardless of whether they reside in the field or in the Antlia cluster. 

We can also make a comparison to samples selected in massive relaxed clusters such as Coma or Virgo \citep[e.g.][]{1995A&A...300L..13B,1997A&A...327..522B}. For Coma, \citet{1997A&A...327..522B} argue that the molecular gas content correlates with SF activity only for high-mass, Milky Way-like ($\sim10^{10.8}$\,M$_{\odot}$) cluster galaxies. Such a relationship is absent for low-mass, lower-metallicity galaxies \citep[e.g.][]{1994A&A...292....1B}, in which the radiation produced by young stars can photo-dissociate the diffuse molecular gas and thus break the expected relationship between CO luminosity and SF, leading to higher $\alpha_\textsc{CO}$ conversion factors. Unlike the results for relaxed clusters, we find that the molecular gas content for Antlia cluster galaxies residing on and below the main sequence strongly correlates with SFR over the entire range of masses probed by our APEX survey, which spans $\sim10^{8.5}-10^{11.5}$\,M$_{\odot}$ (Figure~\ref{fig:CO_luminosity_plots}). \citet{1995A&A...300L..13B} suggest that at lower masses, the amount of molecular gas in cluster galaxies might be underestimated when using $\alpha_\textsc{CO}$ conversion factors derived for more massive galaxies. This would imply that our lower-mass Antlia cluster galaxies would have larger molecular gas reservoirs than field galaxies, pointing to even higher star formation efficiency than in the field. 

From these results we infer that the Antlia cluster represents an intermediate environment between fields and dense clusters, in which the evolving intracluster medium (ICM) may just be starting to affect the member galaxies. We postulate that the ICM in a disturbed cluster environment is not yet dense enough to efficiently strip the molecular gas from the member galaxies as is the case in dense cluster environments, and that this allows the galaxies in the Antlia cluster to retain their reservoirs of molecular gas. However, the process of cluster merging likely disturbs the ICM enough to interact with the member galaxies, inducing the observed high SFRs and possibly beginning to quench the member galaxies. This scenario can explain the high SFRs observed in a number of disturbed cluster environments, as well as the significant molecular gas reservoirs that we have observed in the Antlia cluster. Given that the Antlia cluster represents this intermediate stage in cluster formation, we may consider it is as a useful nearby laboratory to study the processes that likely occur in young clusters residing at higher redshift. For example, \cite{2019PASJ...71...40T} present ALMA CO(3-2) observations of 66 H$\alpha$-selected galaxies in three protoclusters at $z\sim2.5$. They find that galaxies in their sample with stellar masses M$_{\star}<10^{11}$\,M$_{\odot}$ have enhanced molecular gas reservoirs compared to the scaling relations established for field galaxies, while their more massive galaxies (M$_{\star}>10^{11}$\,M$_{\odot}$) have comparable molecular gas reservoirs to galaxies residing in the field. This result suggests that, similarly to our sample of galaxies in the local Antlia cluster, the member galaxies of young clusters in the process of forming at high redshift can also retain their reservoirs of molecular gas.

Resolved studies of galaxies infalling into massive relaxed clusters \citep[e.g.][]{2017MNRAS.466.1382L, 2014ApJ...792...11J, 2018MNRAS.480.2508M} found that molecular CO gas can be slightly stripped along the infall direction, as well as enhanced within sites of intense SF, as traced by H$\alpha$ and far ultraviolet emission. The CO can therefore be modestly enhanced along tails behind and upstream of these galaxies, which may then, in turn, modify the local SF efficiency in the disk \citep{2017MNRAS.466.1382L}. This is in line with simulations which predict that, even in massive clusters, the ram pressure is not enough to strip the molecular gas within the member galaxies, but interactions between these member galaxies and the ICM can disturb and compress the molecular gas which can, in turn, trigger star formation \citep[e.g.][]{2009ApJ...694..789T, 2012MNRAS.422.1609T, 2014MNRAS.443L.114R}. 

In disturbed, lower mass clusters, the ram pressure most likely is not enough to completely remove the molecular gas from infalling galaxies. However, infalling galaxies might experience disturbances in their molecular gas reservoirs caused by interactions with the ICM, which can lead to temporary enhancements in the star formation efficiency. We are capturing Antlia in an active phase of formation, when multiple smaller sub-clusters are undergoing a series of mergers. Therefore, a cluster like Antlia, which is still in the process of formation, might retain a significant number of gas-rich galaxies. Gas-rich galaxies that have already fallen into the cluster prior to the merger between the sub-clusters may be undergoing interactions with the large scale phenomena caused by the cluster formation. Processes such as shocks and turbulence can also temporarily enhance the conversion of atomic into molecular gas which, in turn, will temporarily enhance the star formation activity \citep{2014MNRAS.443L.114R}.

We therefore expect that the galaxies in disturbed cluster environments are either recently quenched, or in the process of becoming quenched. This expectation is supported by Figure \ref{fig:smass_sfr}, where we see that the majority of our sources lie just below, or well below the main sequence, possibly representing populations of galaxies that are in the process of quenching, and are fully quenched.

\subsection{Molecular Gas Fuelling Future Star Formation}

In samples of star forming galaxies residing in the field, as well as in massive, relaxed clusters like Virgo, the depletion time scale of molecular gas correlates with both SFR and stellar mass, but most strongly depends on the specific SFR \citep{2011MNRAS.415...61S, 2016MNRAS.456.4384M}. \citet{2011MNRAS.415...61S} interpret this evolution as evidence for different processes dominating at different stellar masses: morphological (mass) quenching dominates at high masses, while mild starbursts can enhance star formation in lower mass galaxies. 

\begin{figure}
\centering
  \includegraphics[width=\columnwidth]{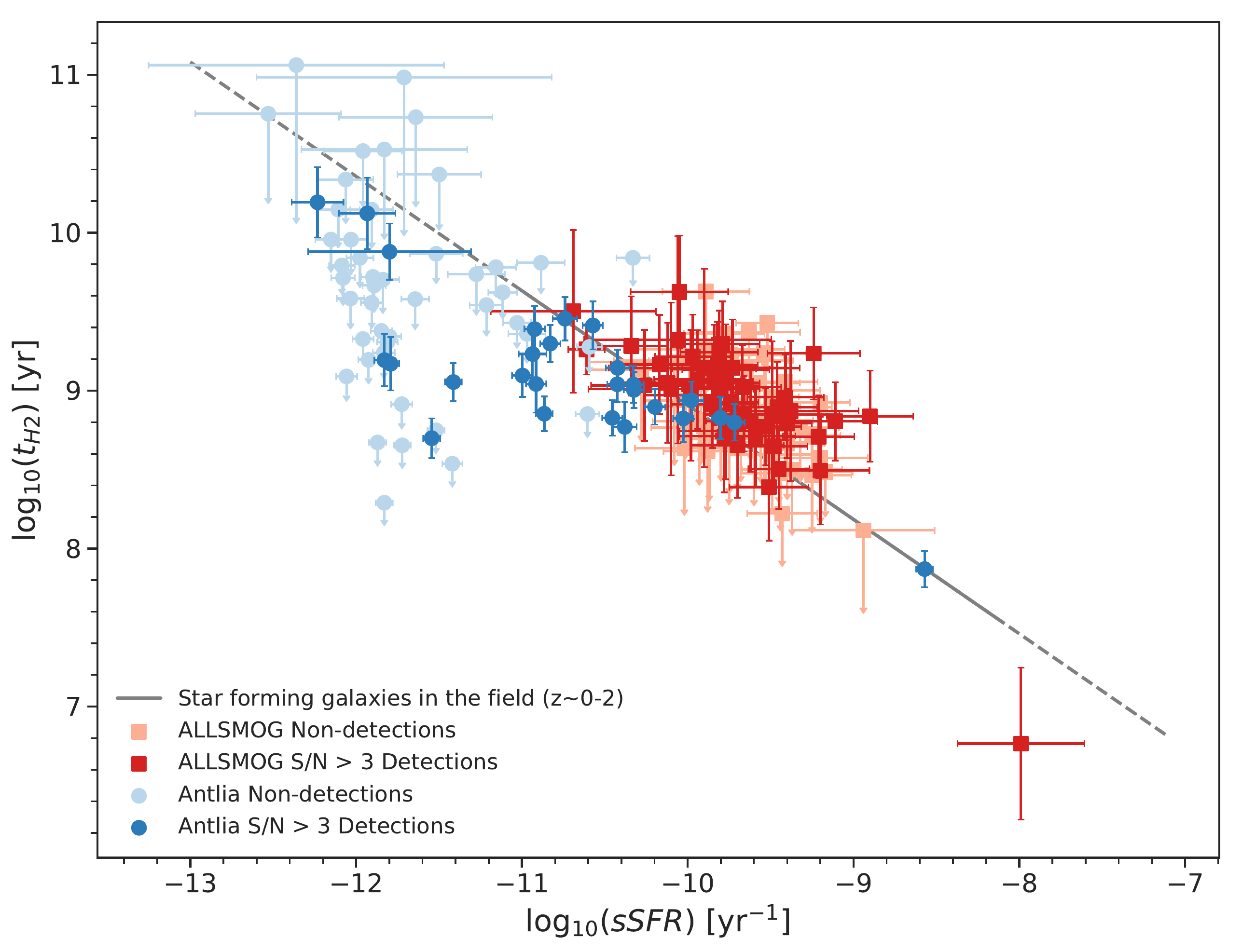}
  \caption{Depletion time scale for molecular gas as function of specific SFR, assuming $\alpha_\textsc{CO}=4.3$\,M$_{\odot}$/(K\,km\,s$^{-1}$\,pc$^2$). We show our cluster galaxies, probing the entire $M_{\star}-$SFR plane, down to a limiting $M_{\star}$ and SFR, as well as main-sequence field galaxies from the ALLSMOG survey \citep{2017A&A...604A..53C}. Both the Antlia cluster and the ALLSMOG field star-forming galaxies follow the relationship derived for star-forming, H{\sc i}-rich field galaxies at $z\sim0-2$ \citep{2011MNRAS.415...61S}. 
  Cluster and field galaxies at $z\sim0.01$ of similar mass, SFR, and consequently specific SFR, have similar depletion timescales, pointing to similar molecular gas reservoirs. Extrapolating the relation to lower values of sSFR, we see a significant spread in the depletion timescales for Antlia galaxies of a given sSFR, primarily driven by our non-detections and sources residing below the main sequence. Cluster galaxies located below the main sequence have on average lower molecular gas reservoirs and shorter depletion times compared to star-forming galaxies of similar masses. This indicates these lower-SFR cluster galaxies have already used up most of their molecular gas and their star formation will be shut down in 100\,Myr to 1\,Gyr.}
  \label{fig:depletion_timescale}
\end{figure}

While our sample of cluster galaxies probes a wider range of masses and SFRs compared to \citet{2011MNRAS.415...61S}, the Antlia main-sequence galaxies fall close to the field relation, well within the scatter of the field star forming galaxies. This indicates that galaxies within the Antlia cluster have similar star formation efficiencies to field galaxies. Galaxies in the Antlia cluster have large molecular gas reservoirs and so, if we consider in the simplest case that gas is depleted only via consumption and conversion into stars, these galaxies will be able to sustain star formation for timescales similar to galaxies residing in the field. This expectation is reflected in Figure \ref{fig:depletion_timescale}, which demonstrates that galaxies in the Antlia cluster with similar sSFRs to those residing in the field also have similar depletion timescales. Moreover, \ref{fig:depletion_timescale} shows that galaxies that lie below the main sequence (i.e.\ lower sSFRs) have lower depletion timescales than star-forming galaxies of  similar masses. The reason for these low depletion timescales is related to Figure \ref{fig:co_smass_sfr}, where we demonstrate that galaxies lying below the main sequence have reduced reservoirs of molecular gas. We conclude that these galaxies are already well on their way to quenching: not only are their SFRs reduced with respect to the main sequence, but also their molecular gas reservoirs are significantly reduced, indicating that star formation in these galaxies will be shut down in 100\,Myr to 1\,Gyr. 

\subsection{Quenching in the Antlia Cluster}
\label{sec:Discussion:Quenching}

Figure \ref{fig:smass_sfr} demonstrates that our sample breaks up into two essentially distinct populations. Just over a third of our sample of 72 galaxies lie slightly below the main-sequence, and follow the trend of the field galaxies residing on the main sequence reasonably well. The vast majority of the remaining galaxies lie well below the main-sequence, and represent a population of galaxies with little ongoing star formation compared to their similar mass counterparts in the field. We therefore believe that, while \cite{2017A&A...604A..53C} uniformly select star forming field galaxies residing on the main-sequence, the galaxies we find in the Antlia cluster are either on their way to becoming quenched, placing them just below the main-sequence, or have already become quenched, placing them far below it. 

In order to investigate this further, we study the morphology of our galaxies (see Appendix \ref{sec:Galaxy_spectra}) to evaluate whether there is any evidence of ongoing quenching. Different methods of quenching can affect the morphology of a galaxy in different ways; hydrodynamical stripping processes could produce a visible tail of material behind the quenching galaxy or a structure reminiscent of a shock on one side, while processes such as harassment can make the galaxy look irregular in morphology. We can also compare these images to the shape of the CO(2-1) emission line in their spectra, allowing us to evaluate whether there is any unusual gas kinematics within them. We also presented in Figure \ref{fig:non_gaussian_distance} the number of our sources which have non-Gaussian CO(2-1) emission profiles as a function of projected cluster-centric distance.  Non-Gaussian or asymmetric CO(2-1) emission line profiles are generally produced by unusual gas kinematics within the galaxy (e.g.\ fast rotating galaxies, hydrodynamical stripping etc.) that could be a result of the galaxy undergoing environmental quenching within the cluster. Figure \ref{fig:non_gaussian_distance} may therefore allow us to determine whether the galaxies that display morphologies or CO(2-1) line profiles reminiscent of quenching galaxies are preferentially located somewhere within the cluster.

From the optical images, there is tentative evidence that a number of our galaxies with non-Gaussian CO(2-1) emission profiles are undergoing quenching. Eleven of our optically selected galaxies are classified as irregular in shape, which could be as a result of galaxy harassment or hydrodynamical stripping. A number of our sources also show asymmetric morphologies and/or features that may imply that they are undergoing some form of stripping. Of our 11 sources that display non-Gaussian CO(2-1) emission line profiles, 5 show some evidence of disturbance in their optical morphology. For the remaining 61 sources (either non-detections or detections that show Gaussian CO(2-1) emission line profiles), a further 8 show evidence of disturbance in their optical morphology. Most notably, J103152 shows evidence of a large shock front on one side of its disc, as does, to a lesser extent, the source J103124. The source J103152 also has a strong double-Gaussian profile, and is the most star-forming galaxy in our sample (see Section \ref{sec:CO_Luminosity_Correlations}), providing tentative evidence that interactions with the ICM can produce disturbed gas kinematics and enhanced SFRs in our sample. J103019 and J102722 are asymmetric, possibly due to material being dragged out of the galaxies via ram pressure stripping. These sources also show unusual profiles in their CO(2-1) emission lines, with the Gaussian profile of J102722 offset by $\sim1000$\,km\,s$^{-1}$ and the CO(2-1) emission line of J103019 being best fit by a double-Gaussian. These galaxies have a range of SFRs and so, if they are becoming quenched, they are likely at different stages in this quenching process. 

Figure \ref{fig:non_gaussian_distance} demonstrates that 5 out of our 11 sources with non-Gaussian CO(2-1) emission line profiles reside at $\sim600$\,kpc from the cluster core, with a further 4 sources with non-Gaussian CO(2-1) profiles residing closer to the cluster core. Although we only have a small number of sources with non-Gaussian CO(2-1) profiles, they tend to reside somewhere between $\sim200$ and $\sim600$\,kpc from the cluster core, where the evolving ICM is likely to be dense enough to begin acting on the member galaxies. Clearly, it is worth noting the small number statistics and the large error bars, but if the sources in our sample with non-Gaussian CO(2-1) profiles do represent a population of galaxies that are undergoing quenching, then this quenching appears to occur primarily between 400 and 800\,kpc from the cluster core.

In order to test whether ram pressure could feasibly cause stripping in our sample, we estimate upper limits on the ram pressure in the Antlia cluster. Generally, the ram pressure acting on a galaxy traversing through a cluster can be approximated as $P_{r} \approx \rho v^{2}$ where $\rho$ is the mass density of the intra-cluster medium (ICM) and $v$ is the velocity of the galaxy relative to the ICM \citep{1972ApJ...176....1G}. We estimate the density of the ICM using the electron number density profile determined by \cite{2016ApJ...829...49W} for the Antlia cluster, converting the number density at a given projected cluster-centric radius into a mass density. We estimate an upper limit on the velocity by assuming that the galaxies are on highly elliptical orbits, in which case their velocities are given by the cluster's escape velocity at the given cluster-centric radius ($v = \sqrt{2GM/r}$) where we use the Antlia cluster virial mass $M \sim 5\times10^{14}$\,M$_{\odot}$. Using this albeit simplistic approximation, we estimate that the ram pressure in the Antlia cluster can reach $\sim10^{-12}$\,ergs\,cm$^{-3}$ within the central $\sim150$\,kpc of the Antlia cluster, falling exponentially to $\sim10^{-15}$\,ergs\,cm$^{-3}$ at $\sim1000$\,kpc from the centre. Simulations by \cite{1999MNRAS.308..947A} estimate that a typical spiral galaxy moving through an ICM of similar density to that of the Coma cluster can feel a ram pressure force of $\sim10^{10}$\,M$_{\odot}$\,(km\,s$^{-1}$)\,kpc$^{-3}$ (corresponding to a ram pressure of $\sim7\times10^{-12}$\,ergs\,cm$^{-3}$), resulting in the stripping of $\sim80\%$ of its diffuse gas mass. It is therefore likely that ram pressure will be strong enough, particularly in the central regions of the Antlia cluster, to strip a significant fraction of the molecular gas from the member galaxies, although it is worth noting that this is a somewhat crude 
estimate  of the ram pressure in the Antlia cluster.

In order to confirm whether the galaxies are undergoing triggered star formation or whether they are on their way to being quenched, we require high-resolution ALMA imaging of our galaxies, which would allow us to determine with great precision the distribution and kinematics of the molecular gas in our sample. This, in turn, would allow us to determine more accurately whether there is ongoing quenching in our galaxies and, if so, which processes are likely responsible.

\section{Conclusions}
\label{sec:Conclusions}

We present the first molecular gas measurements in a complete sample of cluster galaxies. Using the APEX telescope, we carry out CO(2-1) observations of 72 sources in the nearby, disturbed Antlia galaxy cluster. Our survey is unique in its selection of a wide array of galaxy types including both star forming and quenched galaxies selected down to a limiting stellar mass and SFR, as well as its choice of target: a disturbed cluster still in the process of formation. The aim of our survey is to investigate how the molecular gas reservoirs and star formation evolution of galaxies in a merging cluster environment is different from the well-established effect of a relaxed, massive environment.

The 72 galaxies targeted as part of our APEX campaign cover a wide range of stellar mass ($10^{8}$\,M$_{\odot}\lesssim M_{\star} \lesssim 10^{10}$\,M$_{\odot}$) and SFRs (0.0005\,M$_{\odot}$\,yr$^{-1}<\text{SFR}<0.3$\,M$_{\odot}$\,yr$^{-1}$), and so populate a large fraction of the $M_{\star}-$SFR plane. This sample was complemented by a further 20 sources selected with a detection in H{\sc i}, but we infer from these sources that the position of the H{\sc i} detections are highly offset from the CO(2-1) detections, which are co-spatial with the stellar light, and so do not consider these sources in the analysis. 
 
We report a final CO(2-1) detection rate of $\sim37.5\%$ (27/72), with a detection rate for purely star-forming galaxies within 1 dex of the main-sequence of 86\%. We compare the molecular gas properties of the galaxies in our sample to those of a similar sample of field galaxies from the APEX Low-redshift Legacy Survey for Molecular Gas (ALLSMOG, \citealt{2017A&A...604A..53C}). Our conclusions can be summarised as follows: 

\begin{itemize}
    \item While \cite{2017A&A...604A..53C} quite uniformly probe the main sequence for their sample of field galaxies, our sample splits into two distinct populations, one residing just below the main sequence and one residing far below it, possibly corresponding to galaxies that are in the process of quenching and galaxies that are already quenched.
    \item We find that our detection rate increases with cluster centric distance, driven primarily by quiescent elliptical galaxies within 500\,kpc of the cluster core. We further find that our detections cluster towards the higher stellar mass and higher SFR end of our sample, with no observable relation between the CO(2-1) detection rate and redshift.
    \item A number of our galaxies show CO(2-1) line profiles that are non-Gaussian or significantly offset from the expected central position of the CO(2-1) line, reflecting a population with significantly disturbed molecular gas reservoirs. Optical imaging provides tentative evidence that a number of these galaxies are undergoing quenching.
    \item The member galaxies of the Antlia cluster have comparable reservoirs of molecular gas to their counterparts in the field with similar stellar masses and SFRs, contrary to what is seen in virialised clusters. This implies that the member galaxies in disturbed cluster environments are able to hold on to their reservoirs of molecular gas which, in turn, can fuel high SFRs.
\end{itemize}

We therefore conclude that the Antlia cluster may represent the intermediate step between fields containing many star-forming galaxies, and dense clusters in which the majority of their members have been quenched. The gentler ICM of disturbed cluster environments may allow their member galaxies to retain their reservoirs of molecular gas, which can in turn fuel high star formation rates. In this scenario, the evolving ICM may just be beginning to interact with the member galaxies, and hence we expect a number of galaxies in our sample to be undergoing quenching. Further research must be completed in order to confirm that the members of the Antlia cluster are indeed undergoing quenching and to determine the processes that are contributing to this quenching process. While we have proven that galaxies residing in disturbed cluster environments are capable of holding on to their reservoirs of molecular gas, spatially resolved observations of these molecular gas reservoirs with ALMA are required to securely quantify the relationship between the disturbed cluster environment and its member galaxies. On the other hand, a new facility such as the Atacama Large Aperture Submillimeter Telescope (\href{http://atlast-telescope.org}{AtLAST}; see e.g.\ \citealt{Bertoldi2018,Cicone2019}) with a large ($>1^\circ$) field of view capable of surveying the entire cluster field in reasonable integration times at a resolution sufficient to resolve the member galaxies ($\approx5.\!\arcsec4$ at 230 GHz) would present a transformative leap for studies such as that presented here.

\section*{Acknowledgements}

We would like to thank the anonymous referee for their fair and thoughtful comments which have improved the manuscript. We would like to thank Kelley Hess for sharing her HI data and tables and to thank Sandor Kruk for help with the morphological classification of the optical hosts. We thank Fabrizio Arrigoni Battaia for help setting up the observations and Fabrizio Arrigoni Battaia, Palle Moller, Miguel Querejeta, Sthabile Kolwa, Christian Peest, Theresa Falkendal, Jeremy Fensch, and Claudia Cicone for excellent support during the observations. Joseph Cairns and Andra Stroe acknowledge funding from the ESO Science Support Discretionary Funds. Andra Stroe gratefully acknowledges support of a Clay Fellowship. This work is based on observations made with ESO Telescopes at the La Silla Paranal Observatory under programmes ID 0100.A-0316(A) and 0101.A-0681(A), and is partially based on observations made with ESO Telescopes at the La Silla Paranal Observatory under programme IDs 079.B-0480(A) and 093.B-0148(A). Data were obtained from the ESO Science Archive Facility. Some of the data used in this paper were obtained from the Mikulski Archive for Space Telescopes (MAST). STScI is operated by the Association of Universities for Research in Astronomy, Inc., under NASA contract NAS5-26555. This research made use of the Digitized Sky Surveys, produced at the Space Telescope Science Institute under U.S. Government grant NAG W-2166. This research has made use of ``Aladin sky atlas" developed at CDS, Strasbourg Observatory, France. This research has made use of the NASA/IPAC Extragalactic Database (NED), which is operated by the Jet Propulsion Laboratory, California Institute of Technology, under contract with the National Aeronautics and Space Administration.




\bibliographystyle{aasjournal}
\bibliography{references} 




\appendix

\section{Spectra and Optical Images}
\label{sec:Galaxy_spectra}

In Figure~\ref{fig:opticalsources} we show the molecular gas CO(2-1) spectra from APEX, binned to 20\,km\,s$^{-1}$ resolution. We indicate the strength of any detection, using the S/N definition from Section~\ref{sec:defining_detections}. We also show the fit that was used to derive CO properties, which on a case-by-case basis, was a Gaussian, double-Gaussian or a double-horn. In the left panel all sources are shown on the same scale for ease of comparison, while in the middle panel we show the spectra scaled to the peak of the CO(2-1) emission. For each source we also show an RGB image using DSS data. We summarise the properties of the host galaxies in Table~\ref{tab:opticalprop} and the CO(2-1) molecular gas measurements in Table~\ref{tab:opticalCO}.
\startlongtable
\begin{longtable}{lccccccc}
\caption{Properties of the optically-selected sources, including coordinates, velocity, redshift, stellar mass, SFR and optical morphological classification (S-spiral, E-elliptical, I-irregular).}\\
\hline\hline
Source & RA & DEC & v & z & $M_{\star}$ & SFR & Type \\
 & $\mathrm{J2000}$ & $\mathrm{J2000}$ & $\mathrm{km\,s^{-1}}$ &  & $\mathrm{10^9\,M_\odot}$ & $\mathrm{M_{\odot}\,yr^{-1}}$ &  \\
\hline\endhead
J102330 & $10\,23\,30.19$ & $-35\,27\,20.6$ & 4169 & 0.0139 & $2.5\pm0.4$ & $0.061\pm0.002$ & I \\
J102507 & $10\,25\,07.62$ & $-35\,36\,17.9$ & 3472 & 0.0116 & $6.9\pm7.8$ & $0.011\pm0.001$ & S \\
J102622 & $10\,26\,22.22$ & $-34\,57\,48.7$ & 3402 & 0.0113 & $15.5\pm2.1$ & $0.983\pm0.021$ & S \\
J102632 & $10\,26\,32.43$ & $-34\,18\,43.3$ & 3706 & 0.0124 & $16.2\pm2.2$ & $0.025\pm0.001$ & E \\
J102702 & $10\,27\,02.49$ & $-36\,13\,41.1$ & 3122 & 0.0104 & $11.7\pm1.4$ & $1.849\pm0.036$ & S \\
J102720 & $10\,27\,20.43$ & $-35\,16\,27.4$ & 2931 & 0.0098 & $8.5\pm1.2$ & $0.006\pm0.001$ & S \\
J102722 & $10\,27\,22.73$ & $-33\,52\,38.1$ & 2946 & 0.0098 & $5.5\pm0.8$ & $1.052\pm0.022$ & S \\
J102733 & $10\,27\,33.05$ & $-35\,59\,11.4$ & 2722 & 0.0091 & $7.6\pm1.0$ & $0.011\pm0.002$ & S \\
J102757 & $10\,27\,57.38$ & $-35\,49\,18.1$ & 2261 & 0.0075 & $5.1\pm0.8$ & $0.004\pm0.001$ & S \\
J102803 & $10\,28\,03.06$ & $-35\,26\,32.4$ & 3188 & 0.0106 & $4.1\pm0.7$ & $0.154\pm0.005$ & S \\
J102808 & $10\,28\,08.09$ & $-35\,38\,24.9$ & 2986 & 0.0100 & $5.1\pm0.7$ & $0.003\pm0.001$ & S \\
J102816 & $10\,28\,16.03$ & $-35\,32\,01.4$ & 2382 & 0.0079 & $2.3\pm0.4$ & $0.014\pm0.002$ & S \\
J102819 & $10\,28\,19.17$ & $-35\,27\,16.4$ & 2734 & 0.0091 & $16.2\pm2.2$ & $0.015\pm0.002$ & E \\
J10281B & $10\,28\,19.24$ & $-35\,45\,30.6$ & 2519 & 0.0084 & $13.2\pm1.8$ & $0.011\pm0.001$ & S \\
J102823 & $10\,28\,23.97$ & $-35\,31\,46.7$ & 2428 & 0.0081 & $11.0\pm1.5$ & $0.009\pm0.001$ & E \\
J102831 & $10\,28\,31.96$ & $-35\,42\,18.2$ & 2786 & 0.0093 & $14.8\pm2.0$ & $0.021\pm0.001$ & S \\
J102834 & $10\,28\,34.26$ & $-35\,27\,39.1$ & 2890 & 0.0096 & $1.3\pm0.4$ & $0.007\pm0.002$ & I \\
J102847 & $10\,28\,47.12$ & $-35\,39\,29.6$ & 3200 & 0.0107 & $21.9\pm3.0$ & $0.026\pm0.001$ & S \\
J102853 & $10\,28\,53.56$ & $-35\,36\,19.9$ & 2792 & 0.0093 & $154.9\pm17.8$ & $0.250\pm0.006$ & E \\
J102906 & $10\,29\,06.43$ & $-35\,35\,42.4$ & 2416 & 0.0081 & $36.3\pm5.0$ & $0.138\pm0.004$ & E \\
J102911 & $10\,29\,11.05$ & $-35\,41\,16.4$ & 2104 & 0.0070 & $1.6\pm0.3$ & $0.171\pm0.005$ & S \\
J102913 & $10\,29\,13.14$ & $-35\,29\,14.6$ & 1355 & 0.0045 & $5.4\pm0.7$ & $0.005\pm0.001$ & S \\
J102928 & $10\,29\,28.41$ & $-34\,40\,21.8$ & 4093 & 0.0136 & $1.8\pm0.3$ & $0.017\pm0.001$ & U \\
J102930 & $10\,29\,31.00$ & $-35\,15\,35.9$ & 1852 & 0.0062 & $14.5\pm2.0$ & $0.023\pm0.002$ & S \\
J102948 & $10\,29\,48.63$ & $-35\,19\,20.2$ & 3709 & 0.0124 & $45.7\pm6.3$ & $0.040\pm0.002$ & S \\
J102951 & $10\,29\,51.48$ & $-34\,54\,42.1$ & 2549 & 0.0085 & $34.7\pm4.0$ & $0.105\pm0.003$ & S \\
J102953 & $10\,29\,53.04$ & $-35\,22\,30.4$ & 1781 & 0.0059 & $7.6\pm1.0$ & $0.008\pm0.001$ & S \\
J102957 & $10\,29\,57.05$ & $-35\,13\,28.1$ & 3754 & 0.0125 & $66.1\pm7.6$ & $0.089\pm0.003$ & S \\
J10295B & $10\,29\,51.23$ & $-35\,09\,51.1$ & 2609 & 0.0087 & $6.6\pm0.9$ & $0.015\pm0.002$ & E \\
J103000 & $10\,30\,00.65$ & $-35\,19\,31.3$ & 2800 & 0.0093 & $173.8\pm20.0$ & $0.257\pm0.006$ & E \\
J103001 & $10\,30\,01.12$ & $-35\,48\,50.3$ & 1960 & 0.0065 & $2.6\pm0.5$ & $0.003\pm0.001$ & S \\
J103018 & $10\,30\,18.26$ & $-35\,11\,49.1$ & 1768 & 0.0059 & $3.2\pm0.5$ & $0.004\pm0.001$ & E \\
J103019 & $10\,30\,19.51$ & $-34\,24\,18.0$ & 3040 & 0.0101 & $5.1\pm0.7$ & $0.482\pm0.010$ & S \\
J103020 & $10\,30\,20.74$ & $-35\,35\,31.2$ & 2364 & 0.0079 & $0.5\pm0.2$ & $0.007\pm0.001$ & I \\
J103025 & $10\,30\,25.82$ & $-35\,06\,29.1$ & 1781 & 0.0059 & $6.6\pm0.9$ & $0.177\pm0.005$ & S \\
J103026 & $10\,30\,26.48$ & $-35\,21\,34.1$ & 3804 & 0.0127 & $128.8\pm14.8$ & $0.368\pm0.008$ & S \\
J103029 & $10\,30\,29.18$ & $-35\,36\,38.1$ & 2503 & 0.0083 & $61.7\pm7.1$ & $0.117\pm0.003$ & S \\
J10302B & $10\,30\,25.31$ & $-35\,33\,48.2$ & 2140 & 0.0071 & $5.1\pm0.7$ & $0.076\pm0.002$ & S \\
J103044 & $10\,30\,44.95$ & $-35\,21\,32.8$ & 3078 & 0.0103 & $1.1\pm0.3$ & $0.008\pm0.001$ & I \\
J103047 & $10\,30\,47.91$ & $-34\,19\,37.6$ & 2691 & 0.0090 & $1.7\pm0.3$ & $0.013\pm0.001$ & S \\
J103051 & $10\,30\,51.77$ & $-36\,44\,13.2$ & 3161 & 0.0105 & $97.7\pm11.3$ & $1.333\pm0.027$ & S \\
J103059 & $10\,30\,59.61$ & $-34\,33\,46.4$ & 2111 & 0.0070 & $74.1\pm8.5$ & $0.286\pm0.006$ & S \\
J103124 & $10\,31\,24.21$ & $-35\,13\,13.6$ & 2597 & 0.0087 & $10.2\pm1.4$ & $0.385\pm0.009$ & S \\
J103148 & $10\,31\,48.20$ & $-36\,01\,53.6$ & 3168 & 0.0106 & $8.1\pm1.1$ & $0.097\pm0.003$ & S \\
J103152 & $10\,31\,52.11$ & $-34\,51\,13.3$ & 3200 & 0.0107 & $4.0\pm0.5$ & $10.691\pm0.207$ & S \\
J103155 & $10\,31\,55.73$ & $-35\,24\,35.1$ & 2476 & 0.0083 & $39.8\pm5.5$ & $0.059\pm0.002$ & S \\
J103156 & $10\,31\,56.21$ & $-34\,59\,28.9$ & 1981 & 0.0066 & $8.7\pm1.2$ & $0.011\pm0.001$ & E \\
J103158 & $10\,31\,58.46$ & $-35\,11\,53.9$ & 2289 & 0.0076 & $15.8\pm2.2$ & $0.193\pm0.005$ & S \\
J10315B & $10\,31\,52.19$ & $-35\,12\,18.1$ & 2423 & 0.0081 & $9.3\pm1.3$ & $0.012\pm0.001$ & E \\
J103200 & $10\,32\,00.08$ & $-34\,30\,36.7$ & 3571 & 0.0119 & $1.8\pm0.4$ & $0.002\pm0.001$ & S \\
J103208 & $10\,32\,08.44$ & $-34\,40\,29.2$ & 2588 & 0.0086 & $3.5\pm0.6$ & $0.003\pm0.001$ & E \\
J103209 & $10\,32\,09.64$ & $-34\,27\,47.0$ & 2937 & 0.0098 & $1.6\pm0.4$ & $0.017\pm0.002$ & I \\
J103212 & $10\,32\,12.72$ & $-34\,40\,24.0$ & 2129 & 0.0071 & $4.3\pm0.6$ & $0.043\pm0.002$ & E \\
J103214 & $10\,32\,14.47$ & $-35\,15\,33.8$ & 3825 & 0.0127 & $2.1\pm0.3$ & $0.038\pm0.002$ & S \\
J103224 & $10\,32\,24.87$ & $-34\,59\,55.9$ & 3058 & 0.0102 & $5.8\pm0.8$ & $0.271\pm0.007$ & S \\
J103248 & $10\,32\,48.73$ & $-34\,23\,58.1$ & 3023 & 0.0101 & $18.2\pm2.5$ & $0.020\pm0.001$ & S \\
J103259 & $10\,32\,59.55$ & $-34\,53\,10.2$ & 2779 & 0.0093 & $20.4\pm2.8$ & $0.030\pm0.002$ & S \\
J103400 & $10\,34\,00.76$ & $-35\,16\,56.8$ & 2573 & 0.0086 & $9.1\pm1.3$ & $0.432\pm0.009$ & S \\
J103407 & $10\,34\,07.42$ & $-35\,19\,24.0$ & 2754 & 0.0092 & $32.4\pm4.5$ & $0.061\pm0.003$ & S \\
J103408 & $10\,34\,08.69$ & $-34\,38\,01.6$ & 2804 & 0.0093 & $12.9\pm1.8$ & $0.016\pm0.001$ & E \\
J103413 & $10\,34\,13.64$ & $-36\,14\,00.5$ & 3372 & 0.0112 & $1.2\pm0.2$ & $0.030\pm0.002$ & S \\
J103419 & $10\,34\,19.08$ & $-34\,24\,12.1$ & 2644 & 0.0088 & $11.2\pm1.6$ & $0.394\pm0.009$ & S \\
J103445 & $10\,34\,45.04$ & $-35\,28\,14.0$ & 2662 & 0.0089 & $3.0\pm0.6$ & $0.035\pm0.002$ & S \\
J103551 & $10\,35\,51.16$ & $-34\,16\,11.5$ & 3845 & 0.0128 & $3.2\pm0.5$ & $0.135\pm0.004$ & S \\
J103639 & $10\,36\,39.06$ & $-34\,45\,21.8$ & 4061 & 0.0135 & $0.2\pm0.0$ & $0.010\pm0.001$ & S \\
TJ10250 & $10\,25\,05.48$ & $-35\,59\,00.6$ & 3220 & 0.0107 & $2.6\pm0.9$ & $0.008\pm0.001$ & I \\
TJ10274 & $10\,27\,47.07$ & $-34\,31\,56.3$ & 3242 & 0.0108 & $0.3\pm0.1$ & $0.000\pm0.001$ & I \\
TJ10283 & $10\,28\,31.27$ & $-35\,40\,36.2$ & 3608 & 0.0120 & $0.4\pm0.2$ & $0.001\pm0.001$ & I \\
TJ10290 & $10\,29\,01.94$ & $-35\,33\,58.4$ & 2721 & 0.0091 & $0.7\pm0.4$ & $0.001\pm0.001$ & I \\
TJ10300 & $10\,30\,06.12$ & $-35\,23\,22.5$ & 3593 & 0.0120 & $1.1\pm0.5$ & $0.000\pm0.001$ & I \\
TJ10314 & $10\,31\,49.57$ & $-35\,12\,20.4$ & 2423 & 0.0081 & $0.6\pm0.2$ & $0.002\pm0.001$ & I \\
TJ10331 & $10\,33\,16.22$ & $-34\,31\,09.7$ & 1896 & 0.0063 & $3.4\pm0.5$ & $0.001\pm0.001$ & E \\
\hline
\label{tab:opticalprop}
\end{longtable}
\clearpage

\startlongtable
\begin{longtable}{lccccc}
\caption{List of molecular gas properties for the optically-selected sources, including velocity integrated CO line flux, velocity of the CO line, in the restframe of the optical galaxy redshift, CO(1-0) luminosity and molecular gas mass.}\\
\hline\hline
Source & $\int S_{\rm CO(2-1)}{\rm d}v$ & $v_{\rm CO(2-1)}$ & $L_{\rm CO(1-0)}$ & $M_{\rm mol}$ & Fit Type \\
 & $\mathrm{Jy\,km\,s^{-1}}$ & $\mathrm{km\,s^{-1}}$ & $\mathrm{10^6\,K\,km\,s^{-1}\,pc^{2}}$ & $\mathrm{10^8\,M_\odot}$ &  \\
\hline\endhead
J102330 & $<7.6$ & --- & $<10.1$ & $<0.43$ & --- \\
J102507 & $14.3\pm4.5$ & $-276\pm28$ & $19.4\pm7.8$ & $0.83\pm0.33$ & Gaussian \\
J102622 & $132.5\pm8.4$ & $-13\pm2$ & $180.1\pm46.4$ & $7.74\pm2.00$ & Double Gaussian \\
J102632 & $<8.8$ & --- & $<11.8$ & $<0.51$ & --- \\
J102702 & $212.9\pm37.1$ & $1\pm8$ & $288.8\pm88.0$ & $12.42\pm3.79$ & Double Gaussian \\
J102720 & $<9.5$ & --- & $<12.6$ & $<0.54$ & --- \\
J102722 & $112.9\pm10.2$ & $915\pm9$ & $153.9\pm40.9$ & $6.62\pm1.77$ & Gaussian Offset \\
J102733 & $<9.7$ & --- & $<12.9$ & $<0.55$ & --- \\
J102757 & $<9.8$ & --- & $<13.0$ & $<0.56$ & --- \\
J102803 & $36.7\pm3.0$ & $22\pm3$ & $49.9\pm13.1$ & $2.15\pm0.57$ & Gaussian \\
J102808 & $8.0\pm2.4$ & $-209\pm7$ & $10.9\pm4.3$ & $0.47\pm0.18$ & Gaussian \\
J102816 & $<8.5$ & --- & $<11.3$ & $<0.49$ & --- \\
J102819 & $<10.0$ & --- & $<13.4$ & $<0.57$ & --- \\
J10281B & $<9.9$ & --- & $<13.2$ & $<0.57$ & --- \\
J102823 & $<9.7$ & --- & $<13.0$ & $<0.56$ & --- \\
J102831 & $<8.7$ & --- & $<11.7$ & $<0.50$ & --- \\
J102834 & $<6.6$ & --- & $<8.9$ & $<0.38$ & --- \\
J102847 & $<7.1$ & --- & $<9.5$ & $<0.41$ & --- \\
J102853 & $63.7\pm18.8$ & $-206\pm108$ & $86.1\pm33.3$ & $3.70\pm1.43$ & Gaussian \\
J102906 & $<8.3$ & --- & $<11.1$ & $<0.48$ & --- \\
J102911 & $25.6\pm2.7$ & $51\pm2$ & $34.5\pm9.4$ & $1.48\pm0.40$ & Gaussian \\
J102913 & $<7.9$ & --- & $<10.5$ & $<0.45$ & --- \\
J102928 & $<8.0$ & --- & $<10.6$ & $<0.46$ & --- \\
J102930 & $<8.8$ & --- & $<11.7$ & $<0.51$ & --- \\
J102948 & $<8.6$ & --- & $<11.4$ & $<0.49$ & --- \\
J102951 & $<10.3$ & --- & $<13.7$ & $<0.59$ & --- \\
J102953 & $<9.7$ & --- & $<12.9$ & $<0.55$ & --- \\
J102957 & $<7.3$ & --- & $<9.7$ & $<0.42$ & --- \\
J10295B & $<9.9$ & --- & $<13.2$ & $<0.57$ & --- \\
J103000 & $<8.7$ & --- & $<11.6$ & $<0.50$ & --- \\
J103001 & $6.8\pm2.2$ & $389\pm7$ & $9.2\pm3.7$ & $0.40\pm0.16$ & Gaussian \\
J103018 & $<9.7$ & --- & $<13.0$ & $<0.56$ & --- \\
J103019 & $55.0\pm13.3$ & $-3\pm10$ & $74.6\pm25.9$ & $3.21\pm1.12$ & Double Gaussian \\
J103020 & $<7.9$ & --- & $<10.5$ & $<0.45$ & --- \\
J103025 & $79.2\pm19.2$ & $-30\pm8$ & $106.5\pm37.1$ & $4.58\pm1.60$ & Double Gaussian \\
J103026 & $31.4\pm4.6$ & $-24\pm15$ & $42.8\pm12.4$ & $1.84\pm0.53$ & Double Gaussian \\
J103029 & $<9.2$ & --- & $<12.2$ & $<0.53$ & --- \\
J10302B & $26.0\pm2.7$ & $-42\pm5$ & $35.1\pm9.5$ & $1.51\pm0.41$ & Gaussian \\
J103044 & $<8.4$ & --- & $<11.2$ & $<0.48$ & --- \\
J103047 & $<9.5$ & --- & $<12.7$ & $<0.54$ & --- \\
J103051 & $163.0\pm6.6$ & $-0\pm4$ & $221.2\pm56.0$ & $9.51\pm2.42$ & Gaussian \\
J103059 & $55.9\pm6.3$ & $20\pm0$ & $75.3\pm20.7$ & $3.24\pm0.89$ & Double Horned \\
J103124 & $72.6\pm4.9$ & $-49\pm5$ & $98.1\pm25.4$ & $4.22\pm1.10$ & Gaussian \\
J103148 & $40.8\pm9.1$ & $-43\pm10$ & $55.3\pm18.5$ & $2.38\pm0.80$ & Double Gaussian \\
J103152 & $135.5\pm11.4$ & $193\pm8$ & $184.2\pm48.6$ & $7.92\pm2.10$ & Double Gaussian \\
J103155 & $15.8\pm4.5$ & $195\pm33$ & $21.4\pm8.1$ & $0.92\pm0.35$ & Gaussian \\
J103156 & $<10.0$ & --- & $<13.4$ & $<0.58$ & --- \\
J103158 & $36.6\pm12.2$ & $41\pm9$ & $49.5\pm20.6$ & $2.13\pm0.89$ & Double Gaussian \\
J10315B & $<9.7$ & --- & $<12.9$ & $<0.56$ & --- \\
J103200 & $<11.4$ & --- & $<15.3$ & $<0.66$ & --- \\
J103208 & $<11.3$ & --- & $<15.1$ & $<0.65$ & --- \\
J103209 & $<6.8$ & --- & $<9.1$ & $<0.39$ & --- \\
J103212 & $9.2\pm1.7$ & $-24\pm3$ & $12.5\pm3.9$ & $0.54\pm0.17$ & Gaussian \\
J103214 & $18.5\pm3.4$ & $-11\pm9$ & $25.2\pm7.9$ & $1.08\pm0.34$ & Gaussian \\
J103224 & $50.8\pm5.0$ & $2\pm8$ & $68.9\pm18.5$ & $2.96\pm0.80$ & Gaussian \\
J103248 & $<7.4$ & --- & $<9.9$ & $<0.42$ & --- \\
J103259 & $<9.1$ & --- & $<12.1$ & $<0.52$ & --- \\
J103400 & $75.1\pm6.4$ & $22\pm2$ & $101.5\pm26.8$ & $4.36\pm1.16$ & Double Gaussian \\
J103407 & $<8.7$ & --- & $<11.6$ & $<0.50$ & --- \\
J103408 & $<10.0$ & --- & $<13.4$ & $<0.58$ & --- \\
J103413 & $<9.9$ & --- & $<13.2$ & $<0.57$ & --- \\
J103419 & $45.5\pm3.1$ & $6\pm5$ & $61.5\pm15.9$ & $2.65\pm0.69$ & Gaussian \\
J103445 & $10.3\pm2.5$ & $7\pm11$ & $13.9\pm4.8$ & $0.60\pm0.21$ & Gaussian \\
J103551 & $13.6\pm3.6$ & $-6\pm13$ & $18.5\pm6.8$ & $0.79\pm0.29$ & Gaussian \\
J103639 & $<12.1$ & --- & $<16.1$ & $<0.69$ & --- \\
TJ10250 & $<10.2$ & --- & $<13.7$ & $<0.59$ & --- \\
TJ10274 & $<8.4$ & --- & $<11.2$ & $<0.48$ & --- \\
TJ10283 & $<9.4$ & --- & $<12.5$ & $<0.54$ & --- \\
TJ10290 & $<5.9$ & --- & $<7.8$ & $<0.34$ & --- \\
TJ10300 & $<10.0$ & --- & $<13.4$ & $<0.58$ & --- \\
TJ10314 & $<8.1$ & --- & $<10.9$ & $<0.47$ & --- \\
TJ10331 & $<9.9$ & --- & $<13.2$ & $<0.57$ & --- \\
\hline
\label{tab:opticalCO}
\end{longtable}

\captionsetup[subfigure]{labelformat=empty}
\begin{figure*}
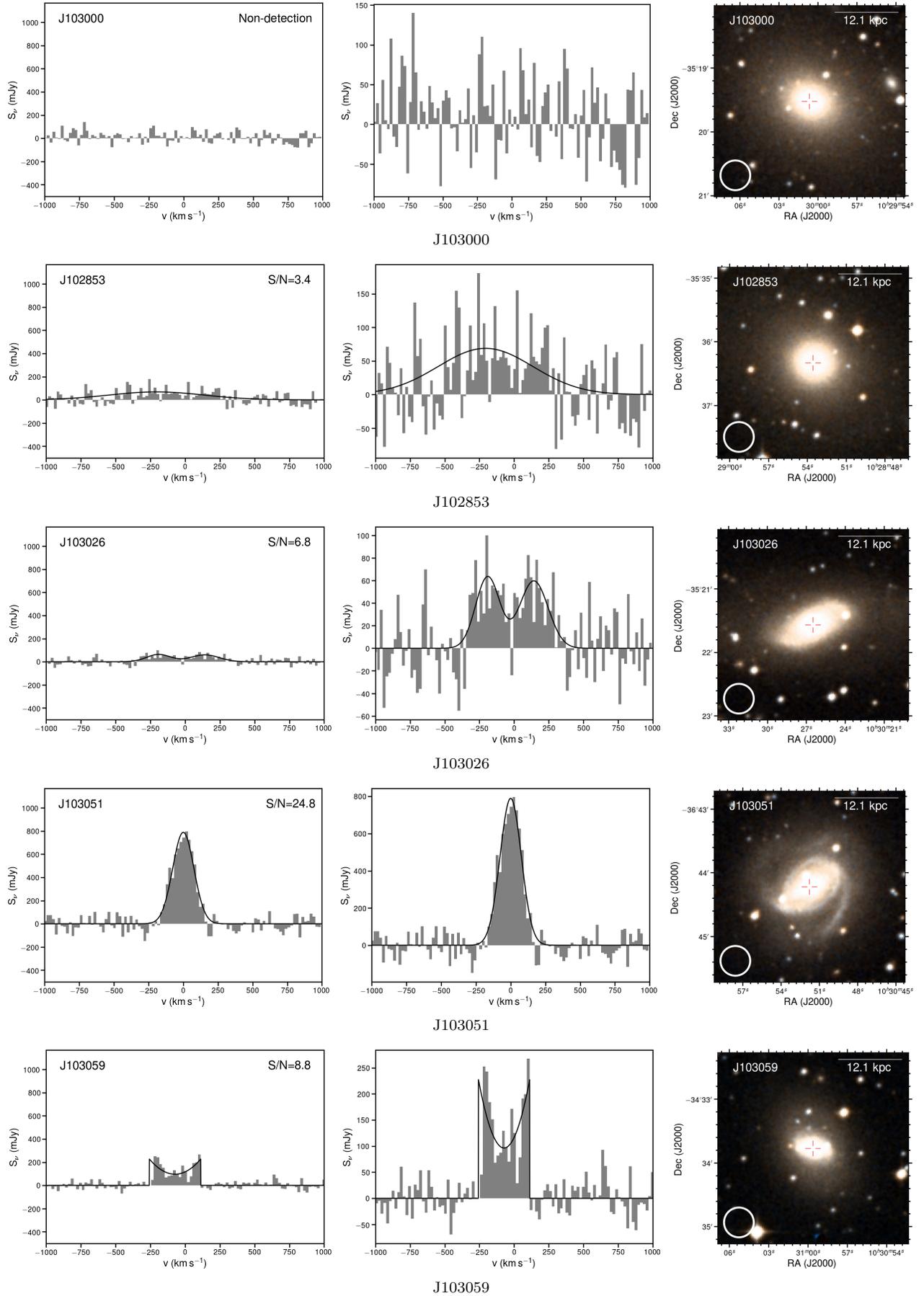

\begin{center}
\foreach \source in {{J103000}, {J102853}, {J103026}, {J103051}, {J103059}}
{\subfloat[\source]{
\includegraphics[height=0.23\textwidth]{Pictures/\source.png}
\includegraphics[height=0.23\textwidth]{Pictures/\source_scaled.png}
\includegraphics[height=0.23\textwidth]{Pictures/\source_rgb.png}
}\vspace{-5pt}\\}
\end{center}
\caption{Sources in the stellar-mass-SFR selection, ordered in descending order of stellar-mass. Left: Observed spectrum in solid gray bars and fitted CO(2-1) Gaussian, double Gaussian or double horned profile in solid black line. All sources on the same scale. Middle: Spectrum scaled to the peak of the CO(2-1) emission. Left: RGB colour-composite from DSS.}
\label{fig:opticalsources}
\end{figure*}

\begin{figure*}
\ContinuedFloat
\begin{center}
\foreach \source in {{J102957}, {J103029}, {J102948}, {J103155}, {J102906}}
{\subfloat[\source]{
\includegraphics[height=0.211\textwidth]{Pictures/\source.png}
\includegraphics[height=0.211\textwidth]{Pictures/\source_scaled.png}
\includegraphics[height=0.211\textwidth]{Pictures/\source_rgb.png}
}\vspace{-5pt}\\}
\end{center}
\caption{Continued.}
\end{figure*}

\begin{figure*}
\ContinuedFloat
\begin{center}
\foreach \source in {{J102951}, {J103407}, {J102847}, {J103259}, {J103248}}
{\subfloat[\source]{
\includegraphics[height=0.211\textwidth]{Pictures/\source.png}
\includegraphics[height=0.211\textwidth]{Pictures/\source_scaled.png}
\includegraphics[height=0.211\textwidth]{Pictures/\source_rgb.png}
}\vspace{-5pt}\\}
\end{center}
\caption{Continued.}
\end{figure*}

\begin{figure*}
\ContinuedFloat
\begin{center}
\foreach \source in {{J102632}, {J102819}, {J103158}, {J102622}, {J102831}}
{\subfloat[\source]{
\includegraphics[height=0.211\textwidth]{Pictures/\source.png}
\includegraphics[height=0.211\textwidth]{Pictures/\source_scaled.png}
\includegraphics[height=0.211\textwidth]{Pictures/\source_rgb.png}
}\vspace{-5pt}\\}
\end{center}
\caption{Continued.}
\end{figure*}

\begin{figure*}
\ContinuedFloat
\begin{center}
\foreach \source in {{J102930}, {J10281B}, {J103408}, {J102702}, {J103419}}
{\subfloat[\source]{
\includegraphics[height=0.211\textwidth]{Pictures/\source.png}
\includegraphics[height=0.211\textwidth]{Pictures/\source_scaled.png}
\includegraphics[height=0.211\textwidth]{Pictures/\source_rgb.png}
}\vspace{-5pt}\\}
\end{center}
\caption{Continued.}
\end{figure*}

\begin{figure*}
\ContinuedFloat
\begin{center}
\foreach \source in { {J102823}, {J103124}, {J10315B}, {J103400}, {J103156}}
{\subfloat[\source]{
\includegraphics[height=0.211\textwidth]{Pictures/\source.png}
\includegraphics[height=0.211\textwidth]{Pictures/\source_scaled.png}
\includegraphics[height=0.211\textwidth]{Pictures/\source_rgb.png}
}\vspace{-5pt}\\}
\end{center}
\caption{Continued.}
\end{figure*}

\begin{figure*}
\ContinuedFloat
\begin{center}
\foreach \source in {{J102720}, {J103148}, {J102733}, {J102953}, {J102507}}
{\subfloat[\source]{
\includegraphics[height=0.211\textwidth]{Pictures/\source.png}
\includegraphics[height=0.211\textwidth]{Pictures/\source_scaled.png}
\includegraphics[height=0.211\textwidth]{Pictures/\source_rgb.png}
}\vspace{-5pt}\\}
\end{center}
\caption{Continued.}
\end{figure*}

\begin{figure*}
\ContinuedFloat
\begin{center}
\foreach \source in {{J103025}, {J10295B}, {J103224}, {J102722}, {J102913}}
{\subfloat[\source]{
\includegraphics[height=0.211\textwidth]{Pictures/\source.png}
\includegraphics[height=0.211\textwidth]{Pictures/\source_scaled.png}
\includegraphics[height=0.211\textwidth]{Pictures/\source_rgb.png}
}\vspace{-5pt}\\}
\end{center}
\caption{Continued.}
\end{figure*}

\begin{figure*}
\ContinuedFloat
\begin{center}
\foreach \source in {{J10302B}, {J103019}, {J102757}, {J102808}, {J103212}}
{\subfloat[\source]{
\includegraphics[height=0.211\textwidth]{Pictures/\source.png}
\includegraphics[height=0.211\textwidth]{Pictures/\source_scaled.png}
\includegraphics[height=0.211\textwidth]{Pictures/\source_rgb.png}
}\vspace{-5pt}\\}
\end{center}
\caption{Continued.}
\end{figure*}

\begin{figure*}
\ContinuedFloat
\begin{center}
\foreach \source in {{J102803}, {J103152}, {J103208}, {TJ10331}, {J103551}}
{\subfloat[\source]{
\includegraphics[height=0.211\textwidth]{Pictures/\source.png}
\includegraphics[height=0.211\textwidth]{Pictures/\source_scaled.png}
\includegraphics[height=0.211\textwidth]{Pictures/\source_rgb.png}

}\vspace{-5pt}\\}
\end{center}
\caption{Continued.}
\end{figure*}

\begin{figure*}
\ContinuedFloat
\begin{center}
\foreach \source in {{J103018}, {J103445}, {TJ10250}, {J103001}, {J102330}}
{\subfloat[\source]{
\includegraphics[height=0.211\textwidth]{Pictures/\source.png}
\includegraphics[height=0.211\textwidth]{Pictures/\source_scaled.png}
\includegraphics[height=0.211\textwidth]{Pictures/\source_rgb.png}
}\vspace{-5pt}\\}
\end{center}
\caption{Continued.}
\end{figure*}

\begin{figure*}
\ContinuedFloat
\begin{center}
\foreach \source in {{J102816}, {J103214}, {J102928}, {J103200}, {J103047}}
{\subfloat[\source]{
\includegraphics[height=0.211\textwidth]{Pictures/\source.png}
\includegraphics[height=0.211\textwidth]{Pictures/\source_scaled.png}
\includegraphics[height=0.211\textwidth]{Pictures/\source_rgb.png}
}\vspace{-5pt}\\}
\end{center}
\caption{Continued.}
\end{figure*}

\begin{figure*}
\ContinuedFloat
\begin{center}
\foreach \source in {{J102911}, {J103209}, {J102834}, {J103413}, {J103044}}
{\subfloat[\source]{
\includegraphics[height=0.211\textwidth]{Pictures/\source.png}
\includegraphics[height=0.211\textwidth]{Pictures/\source_scaled.png}
\includegraphics[height=0.211\textwidth]{Pictures/\source_rgb.png}
}\vspace{-5pt}\\}
\end{center}
\caption{Continued.}
\end{figure*}

\begin{figure*}
\ContinuedFloat
\begin{center}
\foreach \source in {{TJ10300}, {TJ10290}, {TJ10314}, {J103020}, {TJ10283}}
{\subfloat[\source]{
\includegraphics[height=0.211\textwidth]{Pictures/\source.png}
\includegraphics[height=0.211\textwidth]{Pictures/\source_scaled.png}
\includegraphics[height=0.211\textwidth]{Pictures/\source_rgb.png}
}\vspace{-5pt}\\}
\end{center}
\caption{Continued.}
\end{figure*}

\begin{figure*}
\ContinuedFloat
\begin{center}
\foreach \source in {{TJ10274}, {J103639}}
{\subfloat[\source]{
\includegraphics[height=0.211\textwidth]{Pictures/\source.png}
\includegraphics[height=0.211\textwidth]{Pictures/\source_scaled.png}
\includegraphics[height=0.211\textwidth]{Pictures/\source_rgb.png}}\vspace{-5pt}\\}
\end{center}
\caption{Continued.}
\end{figure*}

\section{Molecular gas results from the H{\sc i} selection}
\label{sec:appendix:HI}
Here we discuss the molecular gas observations on a sample of H{\sc i} selected sources in the Antlia cluster. We discuss how the large beam of the H{\sc i} observations resulted in pointing errors in the APEX CO(2-1) observations. Hence no scientific conclusions can be drawn based on these observations.

\subsection{Neutral-gas selection}
Given the blind detection technique of the KAT-7 observations, the H{\sc i} sample from \citet{2015MNRAS.452.1617H} is an ideal complement to the our stellar-mass-SFR selection. The only selection criterion for this sample was a detection in H{\sc i} with a total integrated mass of $M_{HI}>10^{8.2}$\,M$_\odot$. The H{\sc i} sample, consisting of 20 galaxies, is thus selected for its high neutral gas content alone. There are 5 sources in the H{\sc i} sample that have relatively close optical counterparts that were also followed up as part of the stellar-mass-SFR selected sample. Despite being effectively observed twice with sub-mm observations, the pointing was slightly different for the two samples (see Section~\ref{sec:Data_reduction:Observations}).

It is important to note that the sources with (albeit distant, see Section~\ref{sec:Ancillary_Data:HI}) optical counterparts, are on average of lower stellar mass than the stellar-mass sample. The average stellar mass of H{\sc i} rich sources is $\sim10^{8.8}$\,M$_\odot$, reaching down to $\sim10^{7.2}$\,M$_\odot$. The H{\sc i} selection results in a biased redshift distribution that is more broad and systematically offset from the cluster redshift.

\subsection{Observations}

A similar APEX observing strategy as used for the optical sample was also followed for the H{\sc i} selection. An additional 20 H{\sc i}-selected sources were targeted. Unlike the optical selection, the H{\sc i} position was used for those selected through neutral gas (note the caveats through as mentioned in Section~\ref{sec:Ancillary_Data:HI}). Note that 5 sources were observed twice, but with different pointing coordinates as the H{\sc i} and the optical position slightly differed (see also Figures~\ref{fig:opticalsources} and \ref{fig:HIsources}). The tuning frequency was set using the redshift of the H{\sc i} line detection.

The H{\sc i} selected sources were observed between 26-30 Jun 2018, for a total of $\sim21$\,h, of which $\sim15$\,h were spent on science targets. The observing conditions were similar to those attained during the observations of the optically-selected sources (see Figure~\ref{fig:pwv_distribution}) and thus we obtain similar RMS values (as per Figure~\ref{fig:rms_noise}). The CO(2-1) spectra for the H{\sc i} sample can be found in Figure~\ref{fig:HIsources}.

We note that out of the 20 H{\sc i} selected sources, 5 sources were effectively observed twice, by pointing at the optical host galaxy and the H{\sc i} detection (see Figures~\ref{fig:opticalsources} and \ref{fig:HIsources}). 

All 5 sources that were observed based on the optical selection as well as the H{\sc i} follow up are classed as detections in both sets of spectra. Two sources selected in the H{\sc i} follow up that do not have close optical partners (Antlia8 and Antlia23) are classed as detections based on a lower threshold of $S/N>2.5$ (which may be physically motivated, as we discussed in Section \ref{sec:defining_detections} that we likely underestimate our signal-to-noise ratio relative to \citealt{2017A&A...604A..53C}), although they do not appear to be convincing detections.

\subsection{Caveats}
Here we argue that unfortunately no reliable conclusions can be drawn from the APEX observations of the H{\sc i}.

Out of the 20 H{\sc i} selected sources, we obtain 5 detections on sources that have close massive optical counterparts (within $60^{\prime\prime}$). We thus detect 5 sources twice in our survey, as followed up through the H{\sc i} and the optical selection, respectively. We were not able to get any convincing detections on H{\sc i} selected sources with optical counterparts further than $30^{\prime\prime}$. The CO luminosity measured from the H{\sc i} pointing is always lower than the measurement from the optical pointing and drops with distance to the optical host. The fraction of CO recovered in the H{\sc i} pointing drops from 0.65 when the H{\sc i} pointing is offset by $15^{\prime\prime}$ to a mere 0.1 when the pointing is offset by $\sim30^{\prime\prime}$. This is not surprising, given that the size of the APEX beam is $30^{\prime\prime}$. We can conclude that the molecular gas is highly co-spatial with the stellar light. 

It is difficult to ascertain what drives the highly offset H{\sc i} detections. If ram pressure is more efficient at removing H{\sc i} than molecular gas, we would expect to see a larger offset between the optical host and the H{\sc i} detection compared to the CO detection. We would also expect a higher molecular gas detection rate with larger stellar and/or H{\sc i} mass. We thus studied the CO(2-1) detection rate as a function of H{\sc i} mass. We find that there is no obvious dependence of the CO(2-1) detection rate on H{\sc i} mass, although this may be in some part due to the limited number of galaxies for which we have measurements of the H{\sc i} mass, as well as the caveats discussed in Section \ref{sec:Ancillary_Data:HI}. This demonstrates that, for our sample, galaxies with larger H{\sc i} masses are not necessarily more likely to be detected in CO(2-1). We conclude that the spatial offset between the stellar light and the H{\sc i} detection is driving the detections.

Therefore, we conclude that the lack of CO detection in most H{\sc i} selected sources is driven by the large pointing errors. The pointing errors in the APEX measurements were caused by the large uncertainly in the H{\sc i} observations. KAT-7, with which the H{\sc i} data was taken, has a beam of $\sim3^{\prime}$, resulting in a position accuracy of worse than $1^{\prime}$. Given that the resolution of APEX is 30$^{
\prime\prime}$, an average positional error of $1^{\prime}$ can fully explain the flux loss and non-detection of molecular gas. Hence, no reliable interpretation of the molecular gas measurements for our sample of H{\sc i} selected galaxies can be made.

\startlongtable
\begin{longtable}{lccccccc}
\caption{List of H{\sc i}-selected sources, with coordinates of the H{\sc i} detection, H{\sc i} velocity and redshift and H{\sc i} mass. We also list the SFR and stellar mass of most nearby optical counterparts, which in some case is more than 30\,arcsec away. We attribute these mismatches to the large size of the H{\sc i} beam (3\,arcmin) which results in significant pointing errors.)}\\
\hline\hline
Source & RA & DEC & v & z & $M_{\star}$ & SFR & $M_{\rm HI}$ \\
 & $\mathrm{J2000}$ & $\mathrm{J2000}$ & $\mathrm{km\,s^{-1}}$ &  & $\mathrm{10^9\,M_\odot}$ & $\mathrm{M_{\odot}\,yr^{-1}}$ & $\mathrm{10^8\,M_\odot}$ \\
\hline\endhead
Antlia1 & $10\,30\,53.00$ & $-34\,53\,44.0$ & 4108 & 0.0137 & $0.3\pm0.1$ & --- & $8.74\pm0.48$ \\
Antlia12 & $10\,31\,23.00$ & $-35\,13\,10.0$ & 2549 & 0.0085 & $10.2\pm1.4$ & $0.390\pm0.010$ & $4.09\pm0.43$ \\
Antlia19 & $10\,29\,26.00$ & $-35\,00\,57.0$ & 1829 & 0.0061 & $0.1\pm0.0$ & $0.010\pm0.005$ & $3.56\pm0.27$ \\
Antlia20 & $10\,29\,10.00$ & $-35\,41\,20.0$ & 2147 & 0.0072 & $1.6\pm0.3$ & $0.170\pm0.010$ & $4.28\pm0.27$ \\
Antlia22 & $10\,29\,24.00$ & $-34\,40\,50.0$ & 4117 & 0.0137 & $1.8\pm0.3$ & $0.020\pm0.005$ & $3.10\pm0.45$ \\
Antlia23 & $10\,28\,43.00$ & $-34\,41\,54.0$ & 4267 & 0.0142 & $0.0\pm0.0$ & --- & $2.41\pm0.40$ \\
Antlia25 & $10\,27\,00.00$ & $-36\,13\,48.0$ & 3079 & 0.0103 & $11.7\pm1.4$ & $1.850\pm0.040$ & $14.85\pm0.61$ \\
Antlia26 & $10\,30\,13.00$ & $-34\,29\,28.0$ & 4381 & 0.0146 & $0.0\pm0.0$ & $0.000\pm0.001$ & $2.20\pm0.27$ \\
Antlia27 & $10\,32\,22.00$ & $-34\,50\,33.0$ & 4287 & 0.0143 & --- & --- & $5.15\pm0.49$ \\
Antlia28 & $10\,30\,19.00$ & $-34\,24\,31.0$ & 2990 & 0.0100 & $5.1\pm0.7$ & $0.480\pm0.010$ & $3.40\pm0.38$ \\
Antlia30 & $10\,28\,37.00$ & $-36\,26\,00.0$ & 3438 & 0.0115 & --- & --- & $4.80\pm0.11$ \\
Antlia31 & $10\,32\,55.00$ & $-36\,04\,12.0$ & 4145 & 0.0138 & $0.7\pm0.3$ & --- & $1.96\pm0.30$ \\
Antlia32 & $10\,32\,13.00$ & $-34\,32\,10.0$ & 3315 & 0.0111 & $0.0\pm0.0$ & --- & $12.07\pm0.68$ \\
Antlia33 & $10\,31\,00.00$ & $-34\,23\,51.0$ & 3794 & 0.0126 & $0.6\pm0.2$ & $0.010\pm0.005$ & $12.35\pm0.60$ \\
Antlia34 & $10\,30\,09.00$ & $-36\,06\,01.0$ & 2241 & 0.0075 & $0.8\pm0.2$ & $0.010\pm0.005$ & $2.76\pm0.51$ \\
Antlia36 & $10\,26\,55.00$ & $-34\,39\,07.0$ & 3904 & 0.0130 & $0.1\pm0.0$ & $0.010\pm0.005$ & $18.27\pm1.08$ \\
Antlia37 & $10\,26\,57.00$ & $-34\,23\,09.0$ & 3797 & 0.0127 & --- & --- & $16.28\pm0.69$ \\
Antlia4 & $10\,26\,25.00$ & $-34\,57\,58.0$ & 3358 & 0.0112 & $15.5\pm2.1$ & $0.980\pm0.020$ & $6.33\pm0.46$ \\
Antlia6 & $10\,31\,32.00$ & $-35\,32\,06.0$ & 3228 & 0.0108 & --- & --- & $2.10\pm0.30$ \\
Antlia8 & $10\,31\,00.00$ & $-35\,00\,27.0$ & 3142 & 0.0105 & --- & --- & $6.19\pm0.38$ \\
\hline
\end{longtable}

\startlongtable
\begin{longtable}{lccccc}
\caption{List of H{\sc i}-selected sources with CO line properties.}\\
\hline\hline
Source & $\int S_{\rm CO(2-1)}{\rm d}v$ & $v_{\rm CO(2-1)}$ & $L_{\rm CO(1-0)}$ & $M_{\rm mol}$ & Fit Type \\
 & $\mathrm{Jy\,km\,s^{-1}}$ & $\mathrm{km\,s^{-1}}$ & $\mathrm{10^6\,K\,km\,s^{-1}\,pc^{2}}$ & $\mathrm{10^8\,M_\odot}$ &  \\
\hline\endhead
Antlia1 & $<5.3$ & --- & $<7.1$ & $<0.31$ & --- \\
Antlia12 & $29.4\pm5.5$ & $26\pm5$ & $39.7\pm12.4$ & $1.71\pm0.53$ & Double Gaussian \\
Antlia19 & $<8.7$ & --- & $<11.6$ & $<0.50$ & --- \\
Antlia20 & $17.2\pm2.3$ & $7\pm3$ & $23.2\pm6.6$ & $1.00\pm0.28$ & Gaussian \\
Antlia22 & $<7.4$ & --- & $<9.9$ & $<0.42$ & --- \\
Antlia23 & $3.4\pm1.3$ & $41\pm7$ & $4.6\pm2.1$ & $0.20\pm0.09$ & Gaussian \\
Antlia25 & $19.0\pm6.1$ & $111\pm11$ & $25.9\pm10.5$ & $1.11\pm0.45$ & Double Gaussian \\
Antlia26 & $<8.2$ & --- & $<11.0$ & $<0.47$ & --- \\
Antlia27 & $<7.7$ & --- & $<10.2$ & $<0.44$ & --- \\
Antlia28 & $35.3\pm8.1$ & $12\pm9$ & $47.9\pm16.2$ & $2.06\pm0.70$ & Double Gaussian \\
Antlia30 & $<8.2$ & --- & $<11.0$ & $<0.47$ & --- \\
Antlia31 & $<6.7$ & --- & $<9.0$ & $<0.39$ & --- \\
Antlia32 & $<7.3$ & --- & $<9.8$ & $<0.42$ & --- \\
Antlia33 & $<8.4$ & --- & $<11.2$ & $<0.48$ & --- \\
Antlia34 & $<8.0$ & --- & $<10.7$ & $<0.46$ & --- \\
Antlia36 & $<5.4$ & --- & $<7.2$ & $<0.31$ & --- \\
Antlia37 & $<8.0$ & --- & $<10.7$ & $<0.46$ & --- \\
Antlia4 & $20.2\pm5.8$ & $-30\pm4$ & $27.4\pm10.5$ & $1.18\pm0.45$ & Gaussian \\
Antlia6 & $<7.8$ & --- & $<10.4$ & $<0.45$ & --- \\
Antlia8 & $3.4\pm1.3$ & $35\pm12$ & $4.6\pm2.1$ & $0.20\pm0.09$ & Gaussian \\
\hline
\end{longtable}

\begin{figure*}
\begin{center}
\foreach \source in {{Antlia1}, {Antlia4}, {Antlia6}, {Antlia8}, {Antlia12}}
{\subfloat[\source]{
\includegraphics[height=0.211\textwidth]{Pictures/\source.png}
\includegraphics[height=0.211\textwidth]{Pictures/\source_scaled.png}
\includegraphics[height=0.211\textwidth]{Pictures/\source_rgb.png}
}\vspace{-5pt}\\}
\end{center}
\caption{H{\sc i} selected sources, in descending H{\sc i} flux order. Left: Observed sub-mm spectrum in solid gray bars and fitted CO(2-1) Gaussian profile in solid black line. All sources are on the same scale. Solid light gray lines shows the arbitrarily scaled H{\sc i} spectrum.  Middle: Spectrum scaled to the CO(2-1) peak. Left: RGB colour-composite from DSS. }
\label{fig:HIsources}
\end{figure*}

\begin{figure*}
\ContinuedFloat
\begin{center}
\foreach \source in {{Antlia19}, {Antlia20}, {Antlia22}, {Antlia23}, {Antlia25}}
{\subfloat[\source]{
\includegraphics[height=0.211\textwidth]{Pictures/\source.png}
\includegraphics[height=0.211\textwidth]{Pictures/\source_scaled.png}
\includegraphics[height=0.211\textwidth]{Pictures/\source_rgb.png}

}\vspace{-5pt}\\}
\end{center}
\caption{Continued.}
\end{figure*}

\begin{figure*}
\ContinuedFloat
\begin{center}
\foreach \source in {{Antlia26}, {Antlia27}, {Antlia28}, {Antlia30}, {Antlia31}}
{\subfloat[\source]{
\includegraphics[height=0.211\textwidth]{Pictures/\source.png}
\includegraphics[height=0.211\textwidth]{Pictures/\source_scaled.png}
\includegraphics[height=0.211\textwidth]{Pictures/\source_rgb.png}
}\vspace{-5pt}\\}
\end{center}
\caption{Continued.}
\end{figure*}

\begin{figure*}
\ContinuedFloat
\begin{center}
\foreach \source in {{Antlia32}, {Antlia33}, {Antlia34}, {Antlia36}, {Antlia37}}
{\subfloat[\source]{
\includegraphics[height=0.211\textwidth]{Pictures/\source.png}
\includegraphics[height=0.211\textwidth]{Pictures/\source_scaled.png}
\includegraphics[height=0.211\textwidth]{Pictures/\source_rgb.png}
}\vspace{-5pt}\\}
\end{center}
\caption{Continued.}
\end{figure*}

\startlongtable
\begin{longtable}{lcccccc}
\caption{List of H{\sc i} selected sources with optical counterparts within $\sim30^{\prime\prime}$. Note that the CO luminosity is always lower when pointing at the H{\sc I} coordinates, rather than the optical position. The fraction of recovered CO flux is inversely proportional to the distance between the H{\sc I} and the optical position, suggesting that the large H{\sc I} beam causes large pointing errors.}\\
\hline\hline
Source & H{\sc I} name & RA & DEC & $L_{\rm CO(1-0)}$ & $L_{\rm CO(1-0)}$ (HI pointing) & Separation \\
 &  & $\mathrm{J2000}$ & $\mathrm{J2000}$ & $\mathrm{10^6\,K\,km\,s^{-1}\,pc^{2}}$ & $\mathrm{10^6\,K\,km\,s^{-1}\,pc^{2}}$ & $\mathrm{{}^{\prime\prime}}$ \\
\hline\endhead
J102622 & Antlia4 & $10\,26\,22.22$ & $-34\,57\,48.7$ & $180.1\pm46.4$ & $27.4\pm10.5$ & 35 \\
J102702 & Antlia25 & $10\,27\,02.49$ & $-36\,13\,41.1$ & $288.8\pm88.0$ & $25.9\pm10.5$ & 31 \\
J102911 & Antlia20 & $10\,29\,11.05$ & $-35\,41\,16.4$ & $34.5\pm9.4$ & $23.2\pm6.6$ & 13 \\
J103019 & Antlia28 & $10\,30\,19.51$ & $-34\,24\,18.0$ & $74.6\pm25.9$ & $47.9\pm16.2$ & 14 \\
J103124 & Antlia12 & $10\,31\,24.21$ & $-35\,13\,13.6$ & $98.1\pm25.4$ & $39.7\pm12.4$ & 15 \\
\end{longtable}



\end{document}